\documentclass[12pt,letterpaper]{article}
\usepackage[utf8]{inputenc}
\usepackage[version=3]{mhchem} 
\usepackage{siunitx} 
\usepackage{graphicx} 
\usepackage{natbib} 
\usepackage{amsmath} 
\usepackage{graphicx}
\usepackage[outdir=./]{epstopdf}
\usepackage{natbib}
\usepackage{multirow}
\usepackage{xcolor}
\usepackage{newtxtext,newtxmath}
\usepackage{graphicx,subfigure}
\usepackage{multirow}
\usepackage{tabulary,booktabs}
\usepackage{lineno}
\usepackage{amsmath}
\usepackage{hyperref}
\usepackage{xspace}
\usepackage{array}

\newcommand{\kepler}{{\em Kepler\/}}
\newcommand{\tess}{{\em TESS\/}}
\newcommand{\dsct}{\mbox{$\delta$~Sct}}
\newcommand{\corot}{{\em CoRoT\/}}

\title{Complex network view for $\delta$ Scuti stars} 
\author{Elham \textsc{Ziaali}$^{1}$, Nasibe \textsc{Alipour}$^{1}$, and Hossein \textsc{Safari}$^{1,2}$\thanks{E-mail: safari@znu.ac.ir}}

\date{$^{1}$Department of  Physics, Faculty of Science, University of Zanjan, University Blvd., Zanjan, 45371-38791, Zanjan, Iran\\%
$^{2}$Observatory, Faculty of Science, University of Zanjan, University Blvd., Zanjan, 45371-38791, Zanjan, Iran\\%
\today
} 

\begin{document}

\maketitle
\section{Abstract} 
Extraction of characteristics of the complex light curve of pulsating stars is essential for stellar physics. We investigate the complex network (natural and horizontal visibility graphs) properties of the \dsct\ stars light curves observed by \tess. We find that the average shortest path length of \dsct\ light curves is a linear function of the logarithm of the network sizes, indicating the small-world and non-random properties. The small-world property signifies the connection of significant peaks of the light curve with small nearing peaks and other important peaks. The lognormal behavior of nodes' degree is evidence for non-random processes for stellar pulsations. This may be rooted in the different mechanisms of stellar dynamics, such as rotation, excitation of modes, and magnetic activity. The PageRank and nodes' degree distributions of  \dsct\ stars collect in two HADS and non-HADS groups. The lower clustering for HADS than non-HADS indicates a more straightforward light curve (containing one or two independent modes) than a more complex light curve (including various oscillation modes) that might be a signature of surface gravity as an indication of stellar evolution. We show that reducing the window size of a light curve to about 5\% of the original one based on the network ranking preserves most of the star modes information. In this case, we also observe that the amplitude of most natural modes amplifies compared to the noise background in the power spectrum. These results indicate the capability of the network approach for interpreting pulsating stars' light curves. 

\section{Introduction}
Asteroseismology is a fundamental approach for identifying and modeling stellar light curves to determine the internal structure and properties of pulsating stars \citep[e.g.][]{aerts2003, Dalsgaard2004,Baglin2003,Chaplin2009, aerts2010asteroseismology, Handler2013,  basu2017, Bowman2017, Aerts2021, Kurtz2022}. Space telescopes such as \corot\ \citep{Baglin2009}, \kepler\ \citep{Gilliland2010} and \tess\ \citep{Ricker2015} provide the most accurate light curves for astrophysical and exoplanets investigations. The complex network methods for pulsating stars may extract meaningful statistics and hidden characteristics of the system.
In complex network theory, a real-world time series maps to a graph to extract the topological features of the system that do not occur in the simple analysis due to the complex behavior of the system. 

Several analyses have been employed, e.g., on internal structure, mode identification, and classification for \dsct\ stars \citep[e.g.,][]{balona2011, uytterhoeven2011, Chang2013, Kahraman2017, niemczura2015, Pak2018, antoci2019, jayasinghe2020, Tim2020, murphy2020, chaplin2020age, Audenaert2021, bowman2021, Barbara2022, kirmizitas2022}.  
The \dsct\ stars are multi-periodic pulsating stars with intermediate masses. In the HR diagram, the \dsct\ stars are at the transition region between low-mass and high-mass stars. The low-mass stars ($\leq 1M_{\odot}$) have thick convective envelopes, and high-mass stars ($\geq 2M_{\odot}$) have large convective cores and radiative envelopes. These stars have spectral types A0$\text{-}$F5 and pulsate in low-order pressure modes. They are in the lower part of the classical instability strip, within or just above the main sequence stars. Effective temperatures of \dsct\ stars are approximately between 6500\, K and 9500\, K. The \dsct\ stars represent various cepheid-like significant amplitude radial pulsations and the non-radial multi-periodic pulsations within the classical instability strip with dominant pulsation frequencies in the range of 5-80\,d$^{-1}$ \citep[e.g.,][]{Breger2000, Pamyatnykh2000,bowman2016, Michel2017, Balona2018, Bowman+Kurtz2018, Pak2018, Qian2018, ziaali2019,jayasinghe2020, Tim2020, Hasanzadeh2021, barak2022}.

The high amplitude \dsct\ stars (called HADS), are classified in population $\it{I}$ of \dsct\ stars with V band amplitude $\geq 0.3$ mag and also $v\sin i \leq\ 30\ km s^{-1}$. However, the SX Phoenicis type of large amplitude \dsct\ stars is metal-poor population II stars. The frequency spectrum of HADS stars mostly shows only one or two independent modes, which are probably radial \citep{Breger2000, Poretti2003, mcnamara2007, McNamara2011, Balona2012, balona2016, bowman2021kic, Yang2022}. In \dsct\ stars, the well-known $\kappa$ mechanism, which operates in zones of partial ionization of hydrogen and helium, can drive low-order radial and nonradial modes of the low spherical degree to measurable amplitudes (opacity-driven unstable modes). In low temperature \dsct\ stars, near the red edge of the instability strip with substantial outer convection zones, the selection mechanism of modes with observable amplitudes could be affected by induced fluctuations of the turbulent convection. Some studies \citep[e.g.,][]{Houdek1999, Samadi2002, Antoci2011, Antoci2014} suggested the opacity-driven unstable p modes, in which nonlinear processes limit their amplitudes, and intrinsically stable stochastically driven (solar-like) p modes can be excited simultaneously in the same \dsct\ star.  

\citet{deFranciscis2018MNRAS} investigated the fractal property of \dsct\ stars via the rescaled range analysis, multifractal spectra analysis, and coarse-graining spectral analysis. They inferred the scale invariances of light curve of \dsct\ stars may be due to the turbulent. The complex system methods have been widely applied for distinguishing individual and collective features in many scientific fields, such as economics \citep{Souma2003PhyA}, biology \citep{barabasi2004network}, earthquakes \citep{Baiesi-PRE,Pasten2018Chaos,vogel2020measuring}, and Solar physics \citep{Daei2017ApJ, Gheibi2017ApJ, Lotfi2020Chaos,Mohammadi2021JGRA}. The complex network approach could classify features with the same characteristics (e.g.,  the degree distribution, average clustering coefficient, transitivity, and PageRank) and quantify the complexities of dynamic systems. Graph theory is a powerful mathematical tool for exploring the characteristics of complex systems. Time series analysis based on the complex network provides information about feature structure. By mapping the time series into a natural and horizontal visibility graph, we can capture the essential characteristics of the features into distinct individual categories.

Due to small-amplitude nonradial modes in light curves, categorizing the HADS and non-HADS (low amplitude multi-mode \dsct\ stars)  stars is challenging. Amplitude modulation, non-linear driving mechanisms of modes, binarity, and planets' effects are complex features of stars' light curves. Here, we study the characteristics of HADS  and non-HADS stars by mapping the light curves into the graph using the horizontal visibility graph (HVG) and natural visibility graph (NVG) approaches. We study the networks' local and global characteristics (nodes' degree distribution, clustering coefficients, path length, PageRank, etc.) for HADS and non-HADS stars to show the capability of complex network approaches to identifying HADS and non-HADS stars.  

Section \ref{data} explains the \tess\ data set for HADS and non-HADS stars. Section \ref{methods} discuss the methods, including the HVG and NVG algorithms, to map the stellar light curves to the network. Section \ref{results} gives results and discussions. Section \ref{concs} remarks on the important findings of this study. 

\section{Data}\label{data}
\tess\ is a NASA mission and a high-precision photometric instrument \citep{Sullivan2015, Barclay2018} and scans the sky in several sectors for millions of stars at 600-1000 nm bands  \citep{Ricker2015}. \tess\ observed hundreds of thousands of stars with a short cadence of 2 minutes and a long cadence of 30 minutes \citep  {Campante2016,Stassun2018,Feinstein2019} that are collected into MAST\footnote{\url{https://archive.stsci.edu}} in both target pixels and light curve files. Due to fewer data points, the complex network (Section \ref{sec:HVG}) did not form for long cadence \tess\ observation targets. So, we queried the information and short cadence light curves of 33 (Table \ref{tab1}) and 40 (Table \ref{tab2}) HADS and non-HADS stars, respectively, by using the Python Lightkurve package \citep{Lightkurve}.

\section{Methods}\label{methods}

\subsection{Network representation of a stellar light curve}\label{sec:HVG}
A complex network is a helpful approach to studying the characteristics of natural complex systems that evolve via complex dynamics. A complex network extracts a complex system's behavior, composed of individual parts or interacting components. Then show emergent collective characteristics of the system. Natural visibility graph (NVG) and horizontal visibility graph (HVG) are among the several algorithms for building a network for a time series to study the system's characteristics \citep[e.g.,][]{newman2003,luque2009horizontal,newman2010,Gheibi2017ApJ,Daei2017ApJ,Najafi2020ApJ,Lotfi2020Chaos,Mohammadi2021JGRA,taran2022complex}.

\begin{figure}
     \centering
      \begin{subfigure}
         \centering
         \includegraphics[width=8.cm,height=5cm]{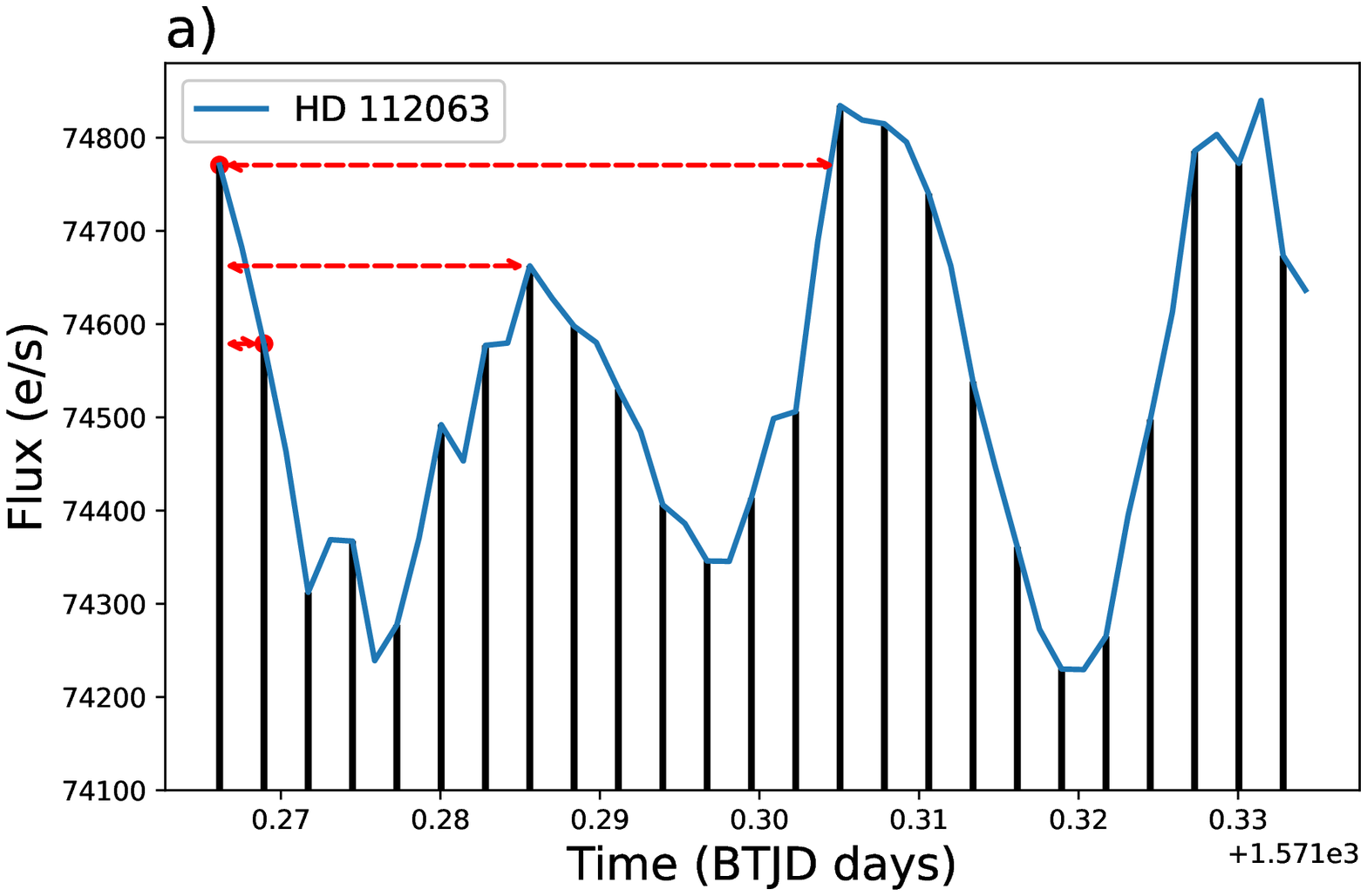}
     \end{subfigure}\\
     \begin{subfigure}
         \centering
         \includegraphics[width=8.cm,height=5cm]{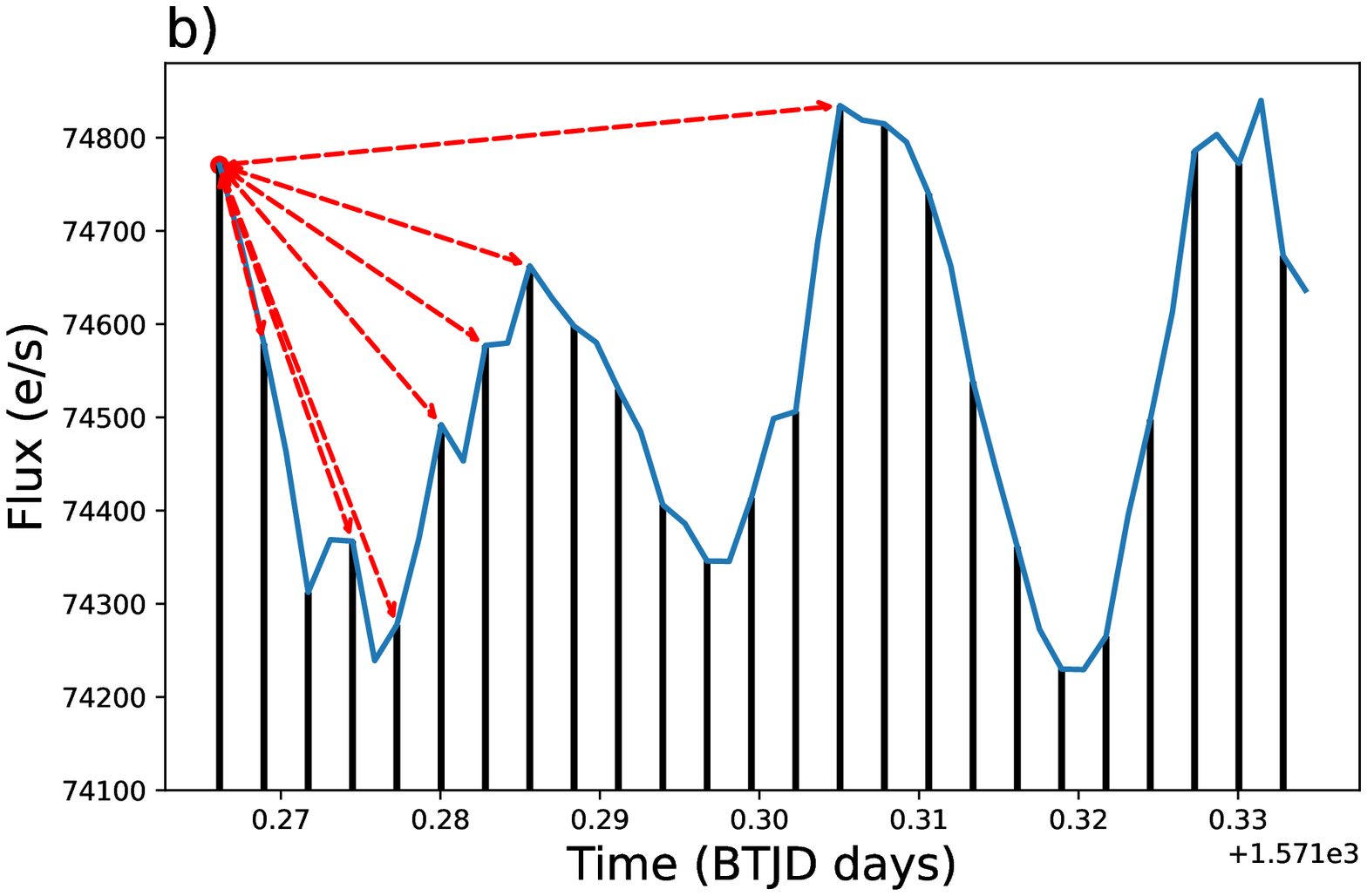}
      \end{subfigure}\\
      \caption{Horizontal visibility (a) and natural visibility (b) for a point of the light curve of HD 112063 \dsct\ star. For HVG and NVG algorithms, the nodes are connected based on Equations (\ref{HVG}) and (\ref{NVG}), respectively.}
\label{fig1}  
\end{figure}

NVG and HVG are algorithms that focus on the interactions of system elements. To map a time series like a stellar light curve to a network, we consider each data point as a node (vertex), and if two elements interact, they are then connected in pairs by lines (edges or links). Figure \ref{fig1} illustrates the connection of nodes (data points) for HD 112063 (TIC 9591460) light curve via the NVG and HVG algorithms. 

For the HVG network for HD 112063  star, node $(t_a, y_a)$ connects to $(t_b, y_b)$ if an arbitrary point $(t_c, y_c)$ such that $t_a<t_c<t_b$ satisfies the condition (Figure \ref{fig1} top panel)
\begin{equation}
     y_a,  y_b > y_c.
    \label{HVG}
\end{equation}

In the NVG algorithm for HD 112063 star, we map the light curve into a network based on the visibility criterion. For two data points $(t_a, y_a)$ and $(t_b, y_b)$ in the light curve have connected nodes in the graph via the criteria:
\begin{equation}
    y_c < y_b +(y_a - y_b)\frac {t_b - t_c}{t_b - t_a}
    \label{NVG}
\end{equation}
in which $(t_c, y_c)$ is an arbitrary data point such that $t_a < t_c < t_b$. 

In the network approach, a set of local and global metrics reflect some particular features of the system. Local metrics describe individual nodes or edges, and global metrics interpret the graph as a whole. We briefly describe the nodes degree, clustering coefficient, shortest path length, and transitivity properties of an HVG network. 

The node's degree of a network is one of the local metrics which measures the centrality. The node's degree is the number of edges linked to the node. Centrality shows the most influential nodes with effective connectivity through the network \citep{Acosta-Tripailao2021Entrp}. 
Local and average clustering coefficients are local metrics in a graph that indicate the node's tendency to cluster together and their neighbors. Transitivity is a global clustering coefficient that determines the density of triangles in a complex network. Transitivity measures the fraction of triples with their third edge served to complete the triangle.
Average shortest-path length is another global metric for complex networks that define the average number of links as the shortest paths for all pairs of nodes. The Google founders developed the PageRank metric that measures the importance of every single node. The assumption is that important nodes have many in-links from essential nodes.

\subsection{Power-law and lognormal distribution fits}\label{fits}
We applied the maximum likelihood estimation in the Bayesian framework \citep{Clauset2009,farhang2018,Alipour2022A&A} to obtain the fit parameters of both power-law and lognormal distributions to data. Also, applying the bootstrapping sampling, we computed the parameters' uncertainty and average (true) value for both the power law and lognormal distributions. We examined a hypothesis test based on K-S (Kolmogorov–Smirnov) statistic. The null hypothesis supposes no significant difference between the nodes' degree distribution and the lognormal/power law model. But, the alternative hypothesis assumes a substantial difference between the nodes' degree distribution and the model. We calculate a p-value to decide whether or not the lognormal/power law distribution hypothesis is plausible for our nodes' degree. A p-value more petite than the threshold of 0.1 refutes the null hypothesis showing that the lognormal/power law distribution is ruled out. We cannot deny the null for a p-value more remarkable than the threshold of 0.1.  

The power-law distribution function is given by 
\begin{equation}
{\rm PDF}(x,x_{\rm min},\alpha)=\frac{\alpha-1}{x_{\rm min}}\left( \frac{x}{x_{\rm min}}\right)^{-\alpha},
\end{equation}
where $x_{\rm min}$ and $\alpha$ are threshold and power index, respectively. This power-law or scale-free property is an essential characteristic of a self-oscillatory process that can be considered as the underlying  mechanism of
self-similar, self-organized, or self-organized criticality systems.

The lognormal distribution function is introduced by 
\begin{equation}
{\rm PDF}(x,\mu,\sigma)=\frac{1}{\sigma\sqrt{2\pi}}\exp\left(-\frac{(\log x -\mu)^2}{2\sigma^2}\right),
 \end{equation}
where $\mu$ and $\sigma$ are the scale and shape parameters, respectively. A lognormal distribution relates to a multiplicative mechanism that shows the effect of the system's independent varying parameters \citep{Bazarghan2008A&A,Farhang2022ApJ}.

\section{Results and discussions}\label{results}
In the current work, we analyzed the complex networks of \dsct\ stars observed by \tess. The HADS stars are mostly radial mono-periodic or double-mode pulsators, but the other \dsct\ stars show several oscillations frequencies (multi modes). We provided a list of HADS stars, including 20 mono-periodic and 13 double-mode HADS stars. We also have 40 non-HADS stars, typically multi-periodic pulsators. We mapped the \tess\ stellar light curves of HADS and non-HADS stars into the individual networks via the HVG and NVG algorithms. 

We applied the HVG and NVG algorithms to investigate the characteristics of 33 and 40 HADS and non-HADS stars, respectively. Due to different algorithms of network structure for HVG and NVG, we represent the results for key metrics of each algorithm. Some network properties in both HVG and NVG algorithms for HADS and non-HADS stars show similar behavior, so we presented the results for one of the algorithms. Also, for some metrics without simple interpretable behavior, we ignored the analysis for those metrics.   

\begin{figure}
    \centering
\includegraphics[width=0.7\linewidth]{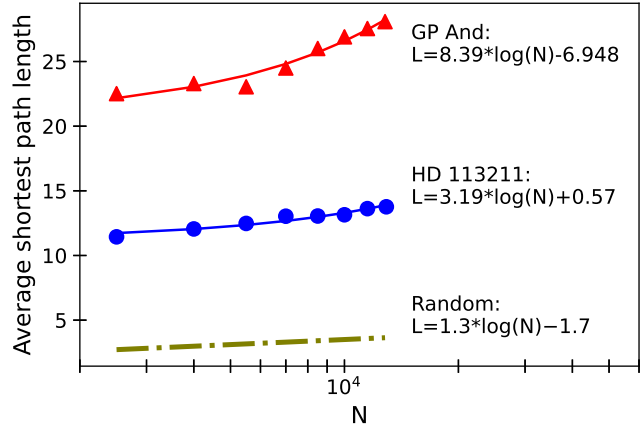}
\caption{The relation between average shortest path length of the HVG network with the network sizes for HD 113211 (blue circles) as a non-HADS star and  GP And star (red triangles) as a HADS star. The fitted line for HD 113211 (blue line) and  GP And star (red line) deviated from the random network (olive dash-dotted line).}
\label{fig2}
\end{figure}

Figure \ref{fig2} represents the average shortest path length for the HVG network of GP And (HADS) and HD 113211 (non-HADS) that compared with the equivalent random network \citep{luque2009horizontal}. As shown in the figure, the average shortest path length network for both stars has deviated from the random network. The linear dependency of the average shortest path length to the logarithm of network size for both cases indicated the small-world behavior of networks. We observed similar behavior for all target stars of Tables \ref{tab1} and \ref{tab2}. The small world behavior for \dsct\ stars network shows that the high peaks at the light curve are connected to several neighboring small peaks and the other high peaks at the light curve. In the context of complex networks, reported a similar behavior for other small-world networks  \citep{Watts-nature1998,Mathias-PRE,Latora-PRL}.  

\begin{figure}
    \centering
\includegraphics[width=13cm,height=8.0cm]{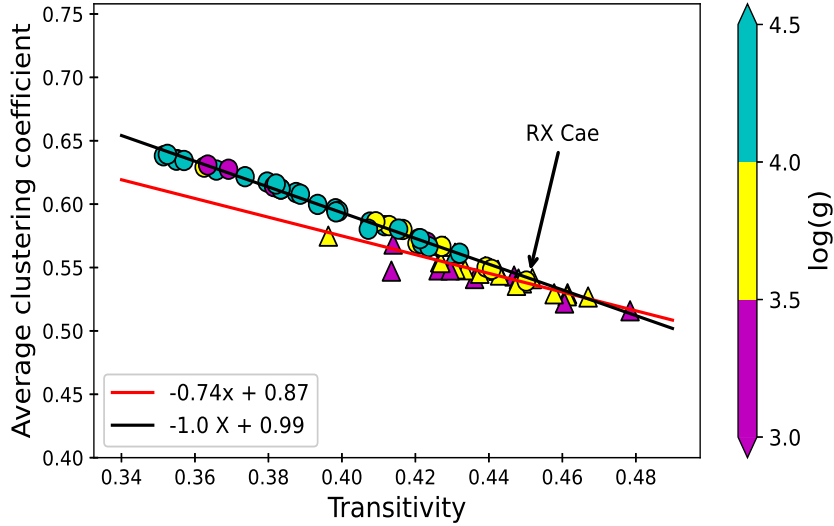}
\caption{Scatter plot of the transitivity versus the average local clustering coefficient for HVG network of non-HADS (circles) and HADS (triangles) stars. The linear fits represent the dependency of the clustering coefficient and transitivity. The age of stars indicated by color bars of surface gravity. The black arrow indicates \dsct\ RX Cae star. }
\label{fig3}
\end{figure}

Figure \ref{fig3} depicts the scattering of the average clustering coefficients and transitivity of HVG networks for HADS (triangles) and non-HADS (circles) stars. The color bar indicates surface gravity, $3 \leq \log g \leq 3.5$ for Terminal-age main sequence (TAMS), $3.5 < \log g \leq 4.0$ for Mid-age main sequence (MAMS), and $4.0 < \log g \leq 4.5$ for Zero-age main sequence (ZAMS) stars \citep{Bowman+Kurtz2018}. The HADS stars have high transitivity and low clustering coefficients, but the other \dsct\ stars have different values. So, the HADS and non-HADS stars are clustered in groups with different slopes (linear fits). The RX Cae (HD 28837) star (indicated by a black arrow) is defined as a HADS candidate. As shown in the figure, the HADS  are mostly MAMS samples that consider the old stars comparing the non-HADS with large surface gravity. The low clustering coefficient indicates the simple light curve (containing one or two independent modes) for HADS compared with the more complicated light curve (including various oscillation modes) for non-HADS. 

\begin{figure*}
     \centering
      \begin{subfigure}
         \centering
         \includegraphics[width=8.5cm,height=2.9cm]{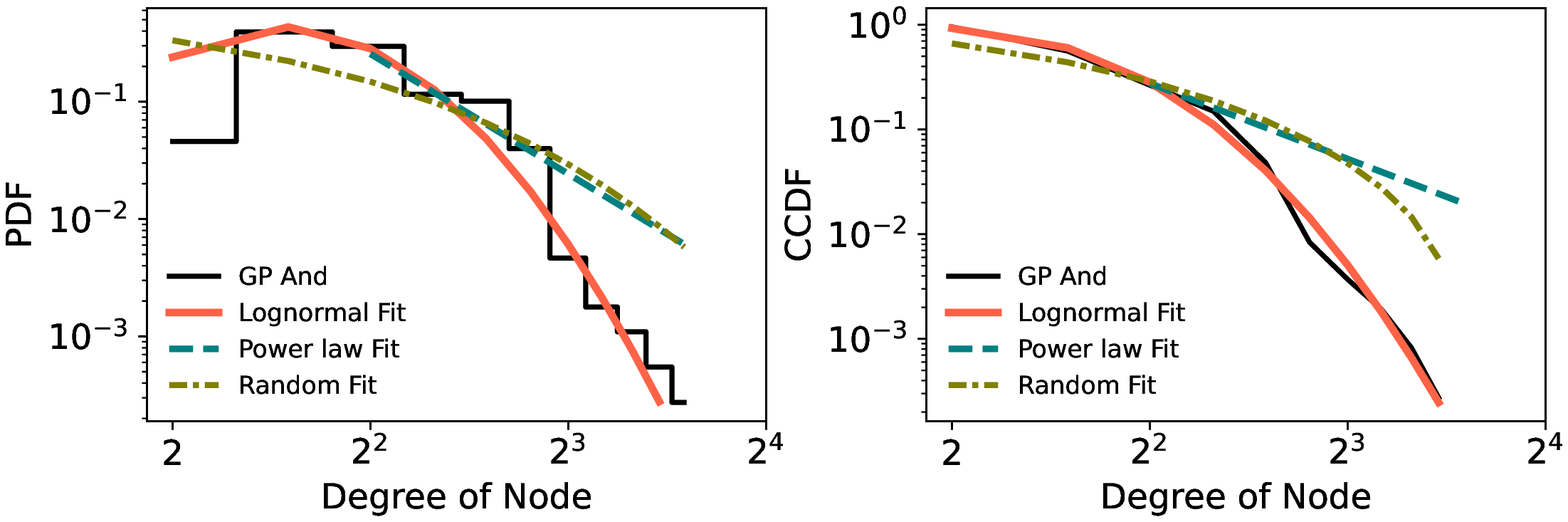}
         \includegraphics[width=8.5cm,height=2.9cm]{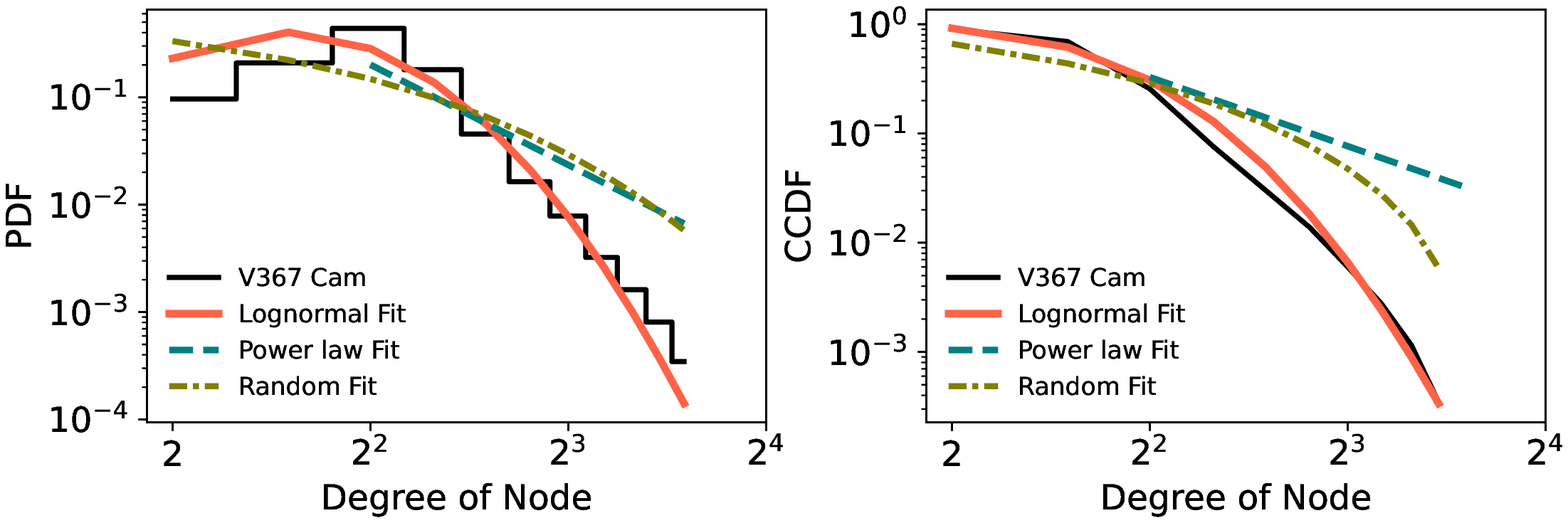}
     \end{subfigure}\\
     \begin{subfigure}
         \centering
         \includegraphics[width=8.5cm,height=2.9cm]{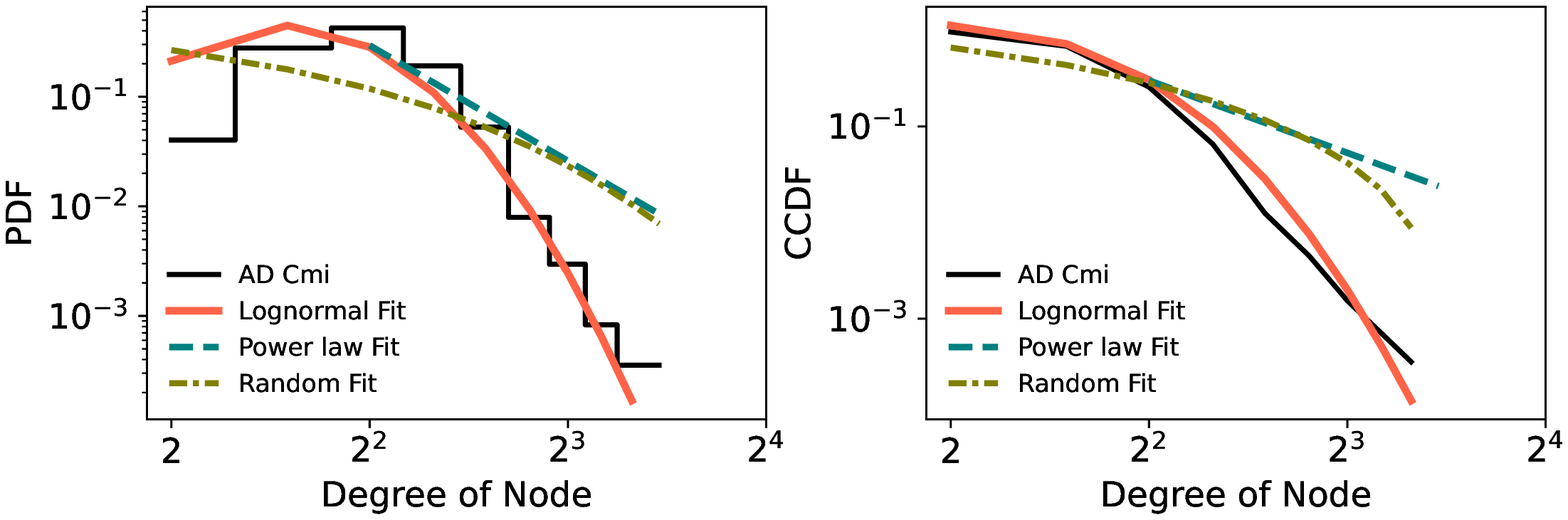}
         \includegraphics[width=8.5cm,height=2.9cm]{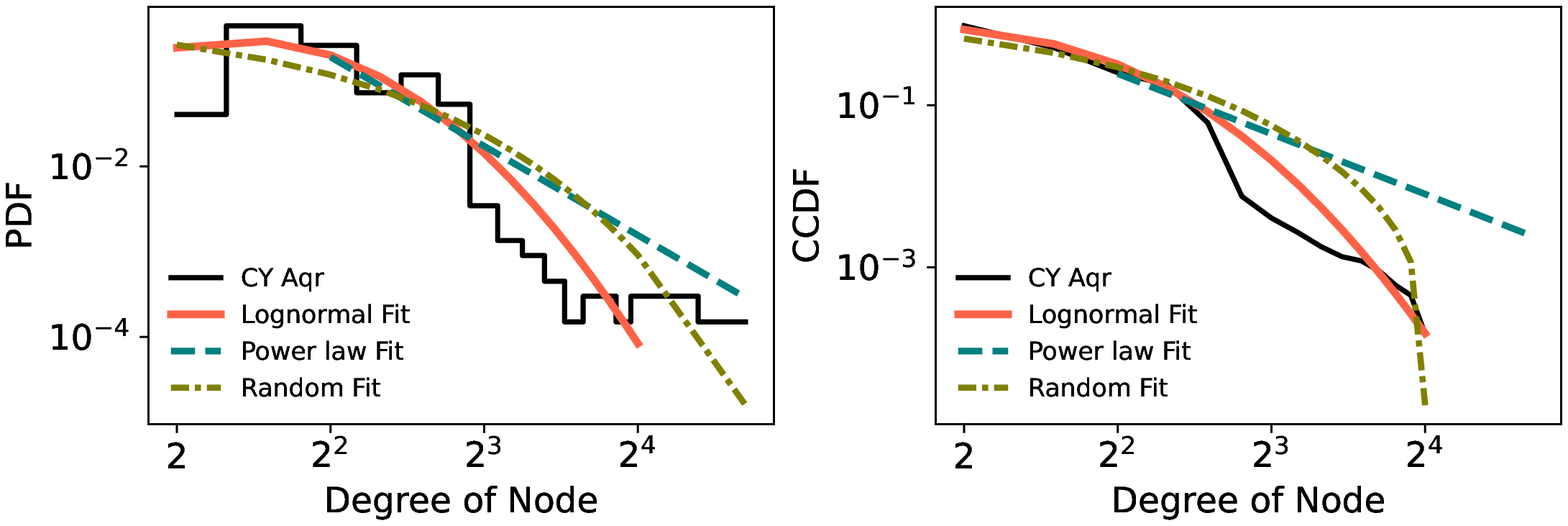}
      \end{subfigure}\\
     \begin{subfigure}
         \centering
         \includegraphics[width=8.5cm,height=2.9cm]{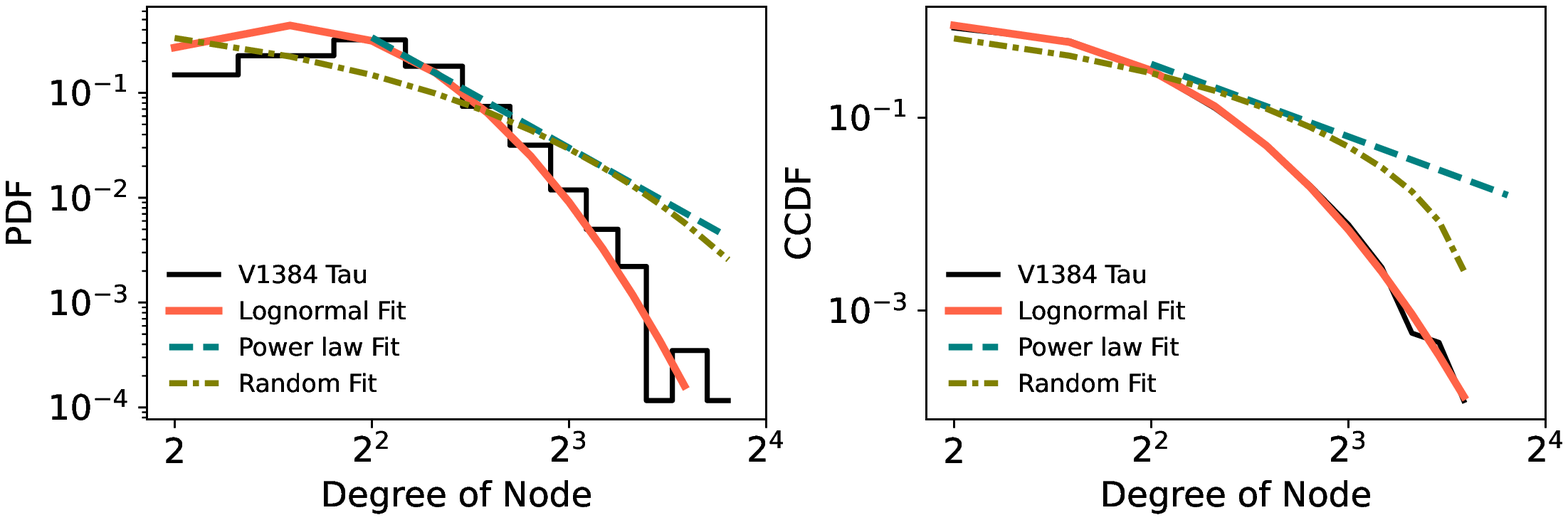}
         \includegraphics[width=8.5cm,height=2.9cm]{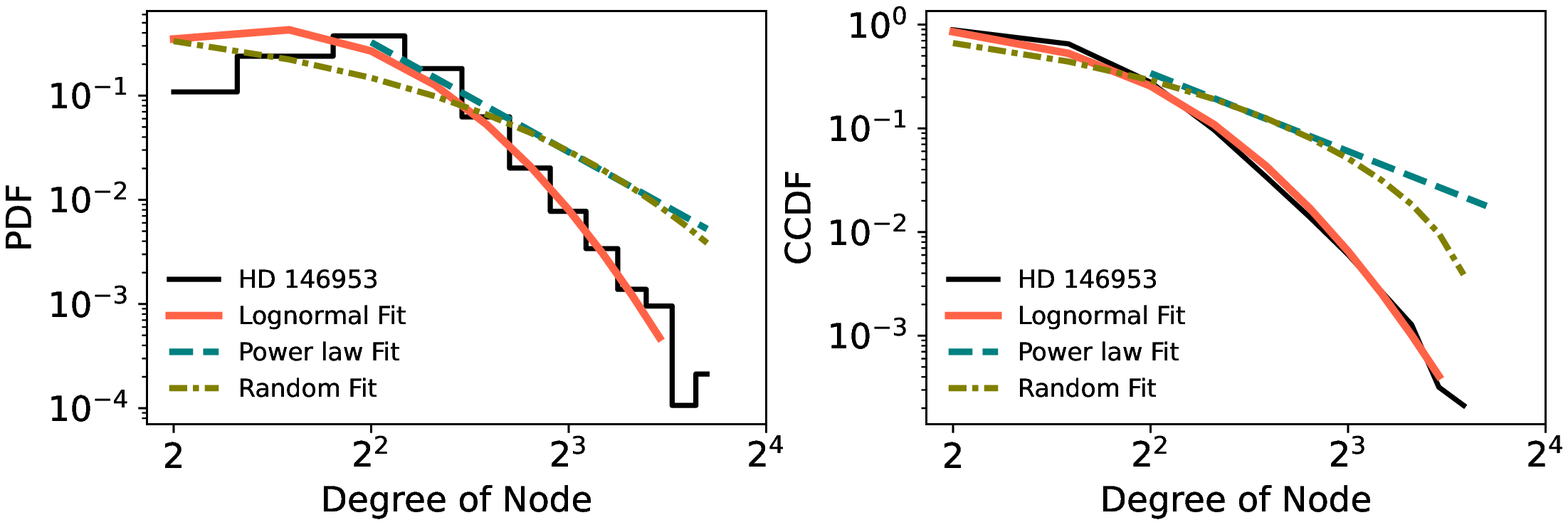}
     \end{subfigure}\\
      \caption{PDF and CCDF for the nodes degree of six HADS stars. Power laws (power law fit: blue dashed lines), lognormal functions (lognormal fit: red line) and random functions (random fit: olive dash-dotted) are fitted to each distribution.}
\label{fig4}  
\end{figure*}

Figure \ref{fig4} shows PDF and CCDF for nodes' degree of the HVG network for six HADS stars. As shown in the figure, the distribution of nodes degree is heavy-tail, so we fitted the power-law and lognormal distribution via the maximum likelihood estimation in the Bayesian framework (Section \ref{fits}).

\begin{figure}
\hspace{2cm}
\includegraphics[width=0.8\linewidth]{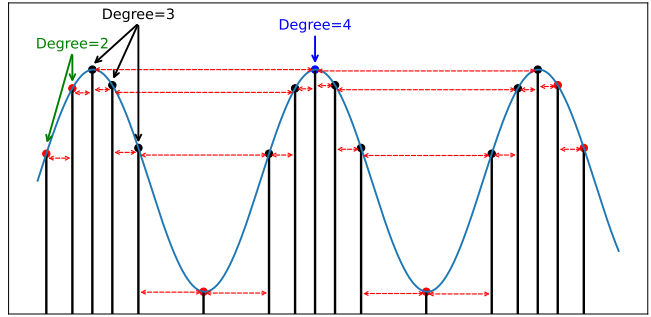}
\caption{Schematic representation of a mono-periodic sinusoidal time series (light curve). The red double head arrows show the links of nodes via the HVG approach. The nodes with the color black, blue, and red have a degree of 3, 4, and 2, respectively. We ignored the network for analyzing two ends of the time series. Nodes with a degree of three are the most frequent.}
\label{fig5}
\end{figure}

We observe that the peak of the nodes' degree distribution (Figure \ref{fig4}) for HADS (mono-periodic: GP And (TIC 436546358), CY Aqr (TIC 422412568), AD Cmi (HD 64191), and V367 Cam (TIC 354872568); double mode: V1384 Tau (TIC 415333069) and HD 146953) are precisely three or four. 
We analyzed the HVG characteristics of a single-mode artificial sinusoidal light curve to address this essential property of HADS stars. Figure \ref{fig5} displays the horizontal visibility required for the HVG network for a part of a mono-periodic artificial time series. We observe that the nodes with three links are more frequent than the others, with the number of connections differing from three. In other words, we follow that the nodes (black points) connect to three other nodes due to the behavior of a sinusoidal mono-periodic artificial time series—this property of a fully sinusoidal time series with various frequencies/periods. However, most mono-periodic stellar light curves have the HVG node distribution peaks of about three or four, similar to the artificial time series. However, the range of degrees of nodes is different for regular artificial time series and mono-periodic stellar light curves. These differences may indicate the diverse nature of light curves of pulsating stars as $\kappa$, stochastic, or other complex generative mechanisms \citep[][]{Houdek1999, Samadi2002, Antoci2011, Antoci2014} for pulsations from a regular sinusoidal oscillation. We find the p-value for both power-law fit and random model \citep{luque2009horizontal} is less than 0.01, which rejects both models for nodes' degree distribution of \dsct\ stars. The rejection of the random model implies that the \dsct\ stars light curve is not random. The p-values for lognormal fits are more than 0.1 (Table \ref{tab1}), which indicates we can not refute the lognormal model for nodes' degree distribution of HADS stars. We obtain the lognormal distribution parameters in the range of 3.05 to 3.86 and 0.2 to 0.6 for scale and shape, respectively. 
The lognormal distribution is an essential feature of systems handled with multiplicative independent varying parameters \citep[][]{mitzenmacher2004,Bazarghan2008A&A, Tajfirouze2012ApJ, tokovinin2014binaries, ruocco2017bibliometric,Alipour2022A&A,Farhang2022ApJ}. The lognormal distribution of the network parameter for \dsct\ stars might originate in the pulsating driving mechanisms. The different mechanisms of stellar dynamics control the pulsating driving mechanisms, including rotation, excitation of modes, and magnetic activity \citep{deFranciscis2018MNRAS}.

\begin{figure*}
     \centering
      \begin{subfigure}
         \centering
         \includegraphics[width=8.5cm,height=2.9cm]{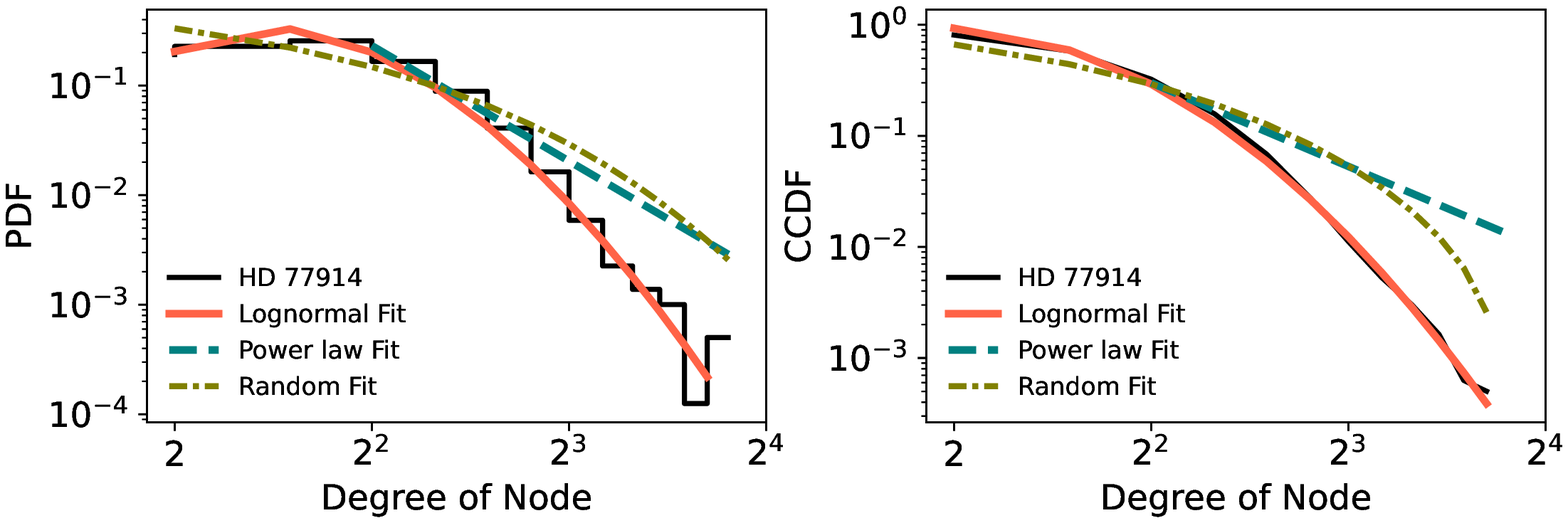}
         \includegraphics[width=8.5cm,height=2.9cm]{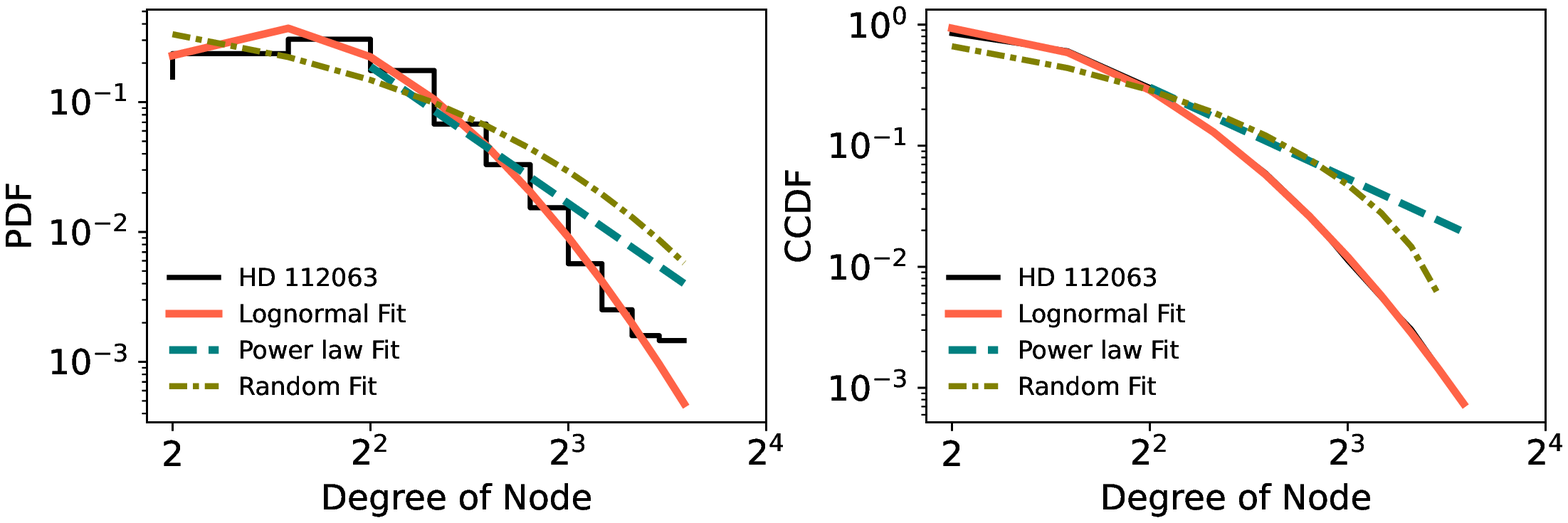}
     \end{subfigure}\\
     \begin{subfigure}
         \centering
         \includegraphics[width=8.5cm,height=2.9cm]{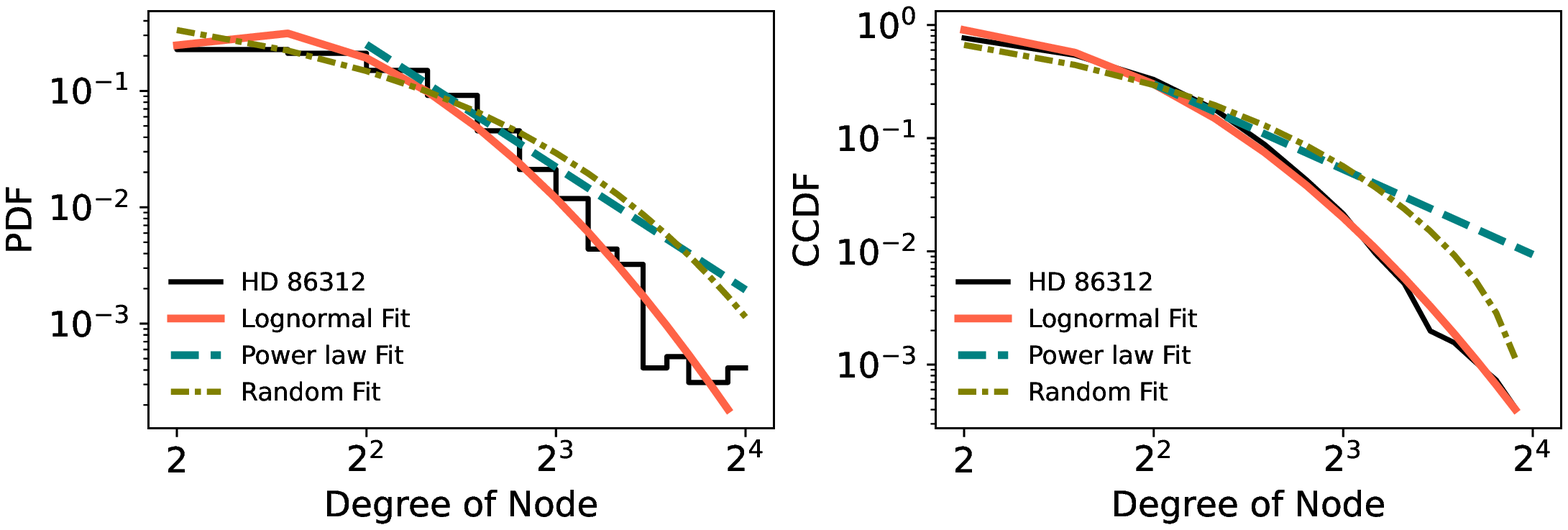}
         \includegraphics[width=8.5cm,height=2.9cm]{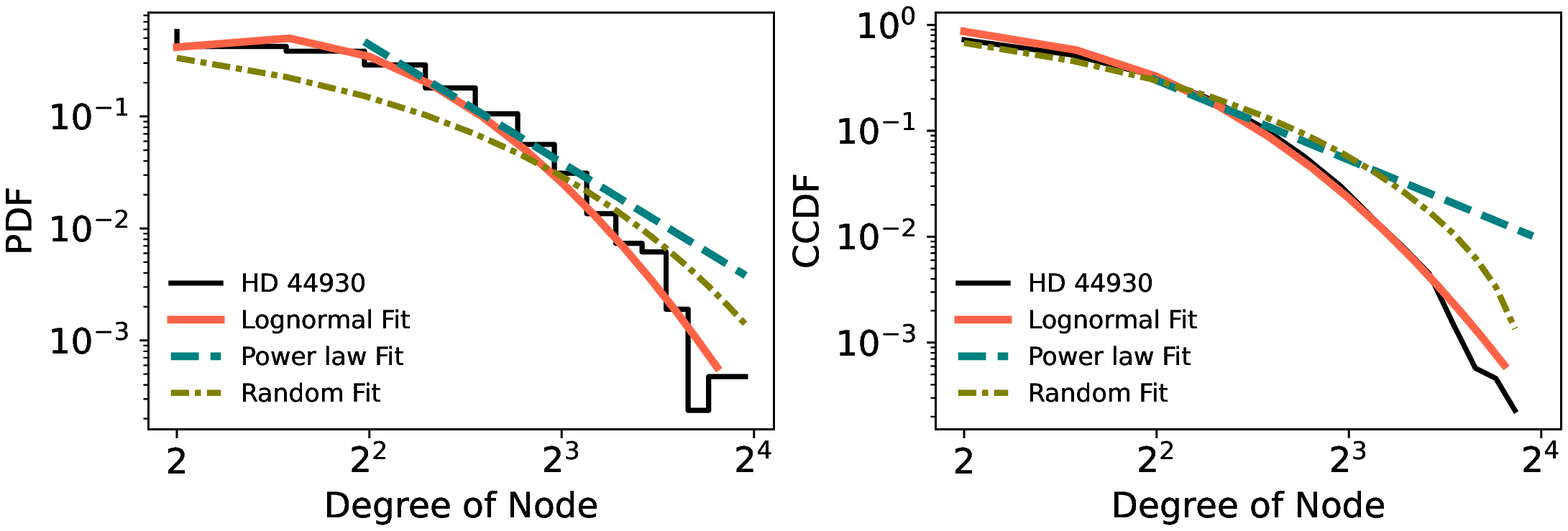}
      \end{subfigure}\\
     \begin{subfigure}
         \centering
         \includegraphics[width=8.5cm,height=2.9cm]{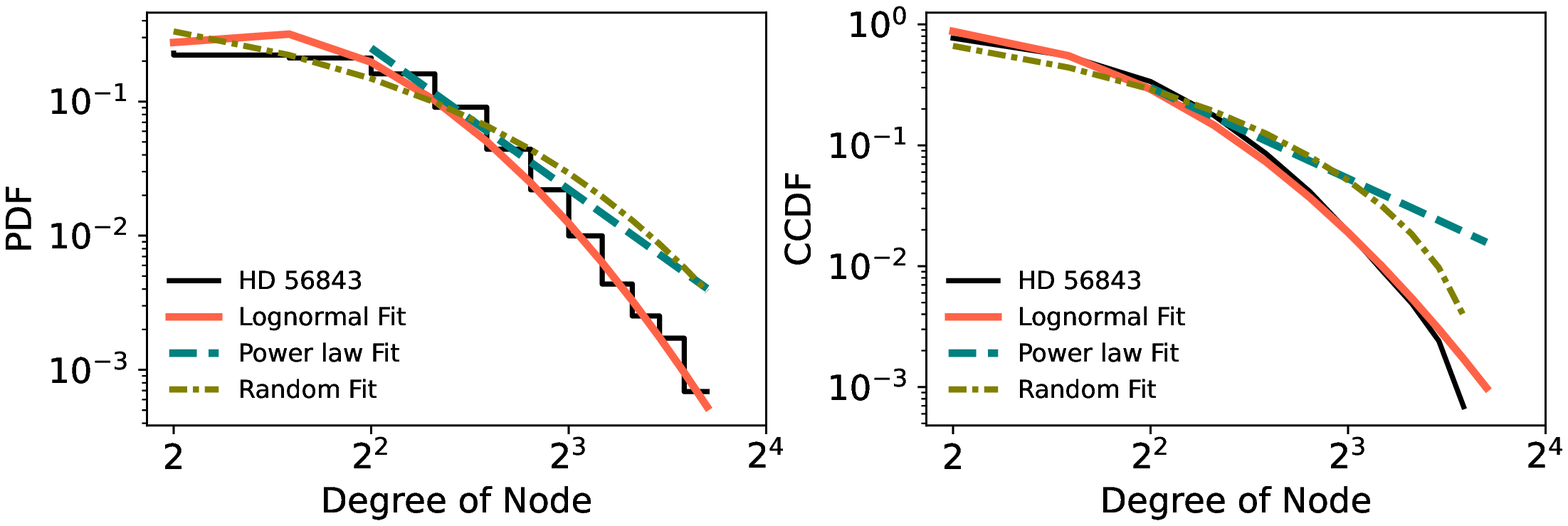}
         \includegraphics[width=8.5cm,height=2.9cm]{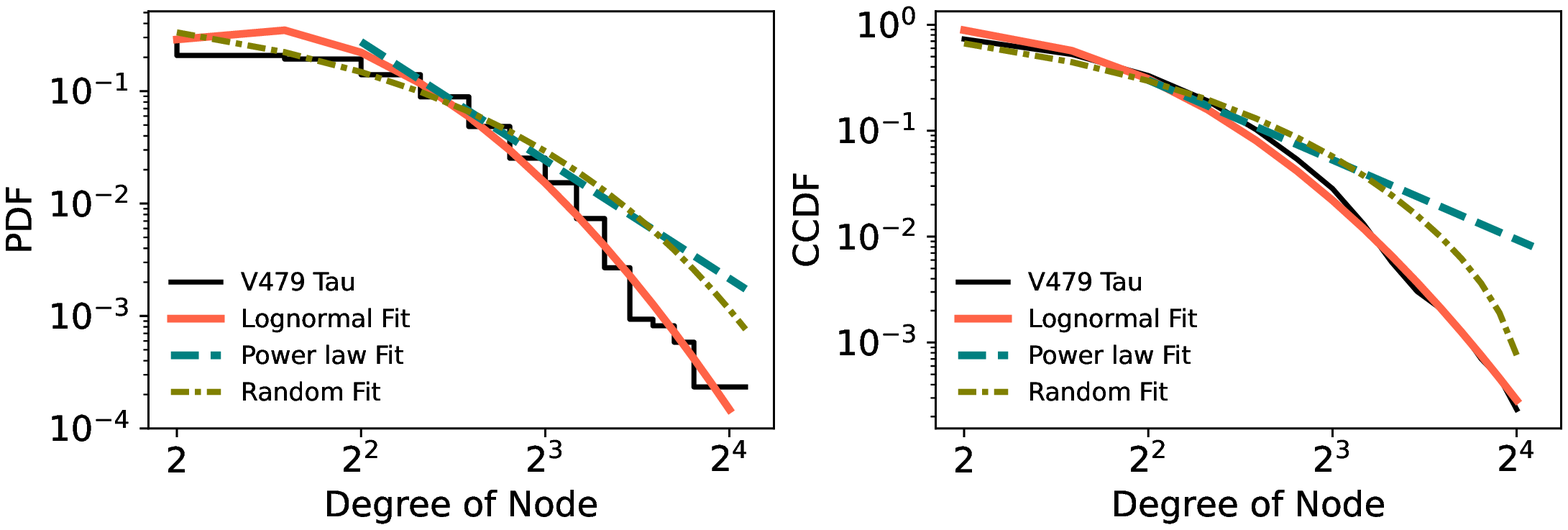}
     \end{subfigure}\\
      \caption{PDF and CCDF for the nodes degree of six non-HADS stars. Power laws (power law fit: blue dashed lines), lognormal functions (lognormal fit: red line) and random functions (random fit: olive dashdot) are fitted to each distribution.}
\label{fig6}  
\end{figure*}

Figure \ref{fig6} shows PDF and CCDF for nodes' degree of the HVG network for six non-HADS stars, HD 77914, HD 112063, HD 86312, HD 44930, HD 56843, and V479 Tau (HD 24550). We obtain the p-values more than 0.1 (Table \ref{tab2}) for lognormal fits, which shows we can not reject the lognormal model for nodes' degree distribution of non-HADS stars. We find the lognormal distribution parameters in the range of 2.4 to 3.9 and 0.33 to 0.6 for scale and shape, respectively. A lognormal distribution is due to a multiplicative process \citep[e.g.,][]{Hubble1934ApJ,Kolmogorov1962JFM,Miller1979ApJS,McBreen1994MNRAS,Mouri2009PhFl}. So, the lognormal behavior of nodes degree for stellar flux network may indicate the flux of \dsct\ stars forming via some multiplicative process. \citet{Pauluhn2007A&A} and \citet{Bazarghan2008A&A} showed that the small-scale solar brightening features distribution follows a lognormal distribution. 

\begin{figure}
      \centering
      \begin{subfigure}
         \centering
         \includegraphics[width=7.5cm,height=4.5cm]{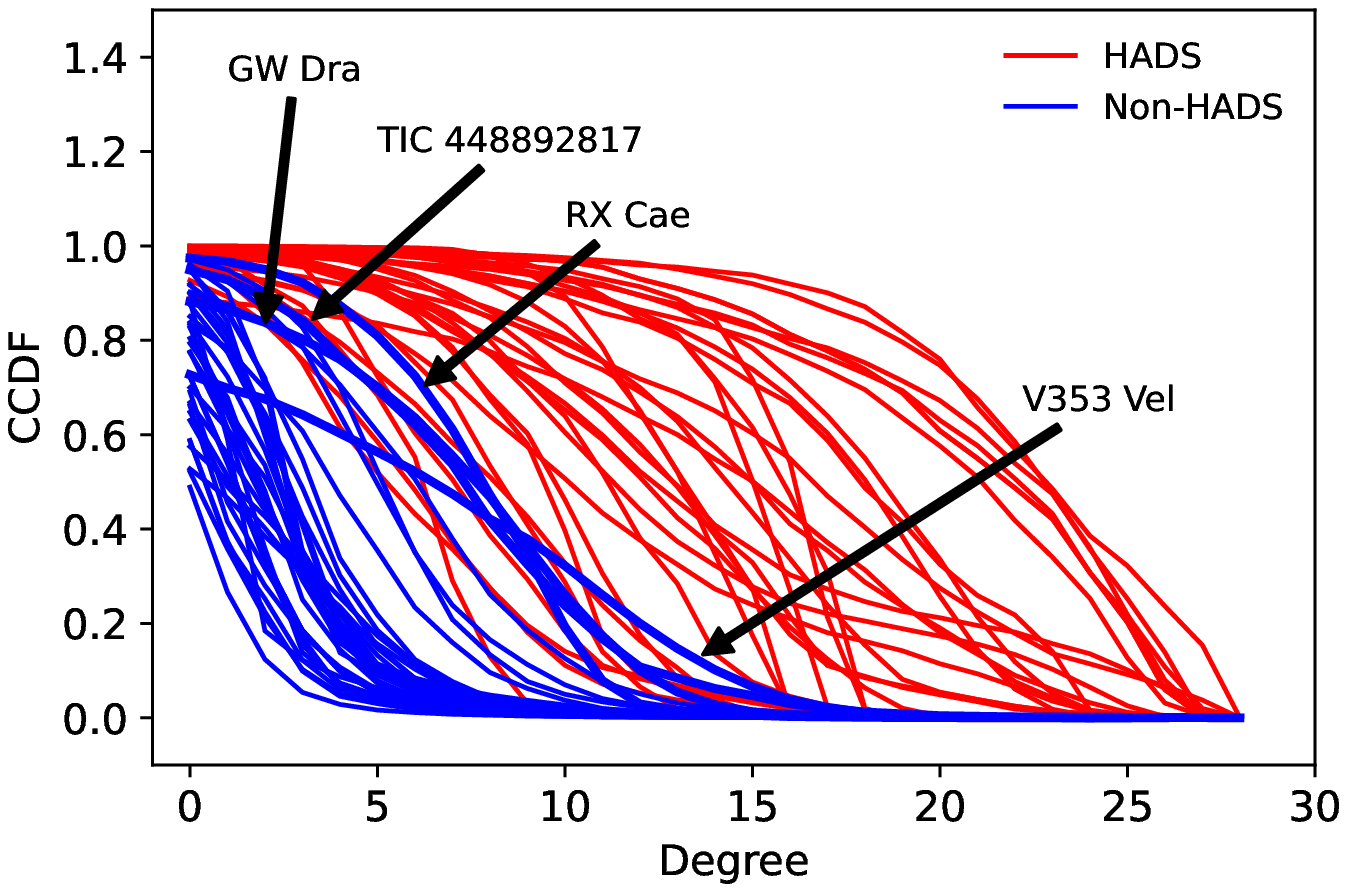}
     \end{subfigure}\\
     \begin{subfigure}
         \centering
         \includegraphics[width=7.5cm,height=4.5cm]{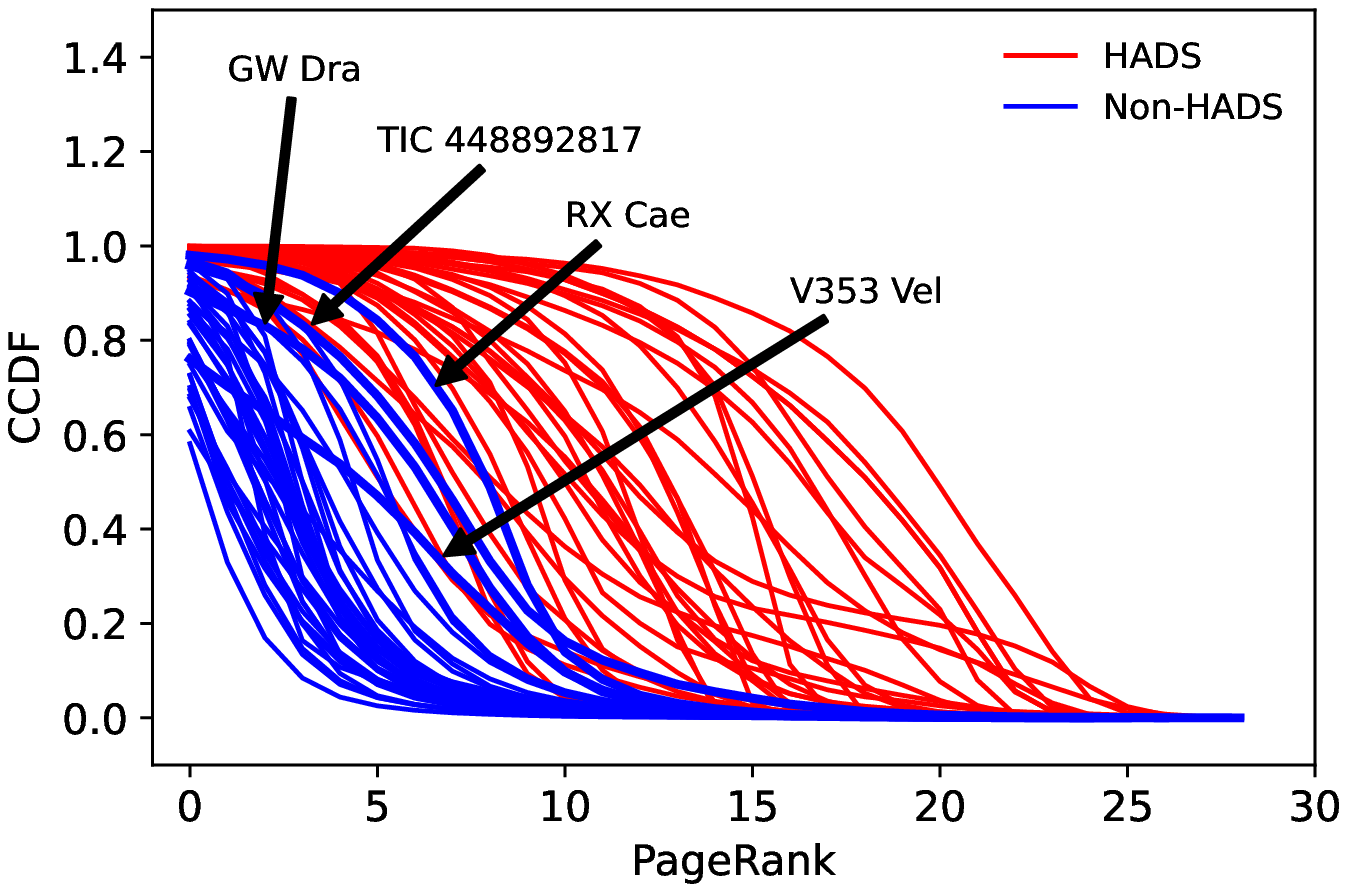}
      \end{subfigure}\\
\caption{CCDF of nodes' degree (top panel) and PageRank (bottom panel) of NVG network for HADS (red lines) and non-HADS (blue lines) stars. The  CCDF of nodes' degree and PageRank for RX Cae, V353 Vel, TIC 448892817, and GW Dra are samples that show different behavior of multi-mode \dsct\ stars.}
\label{fig7}
\end{figure}

Figure \ref{fig7} depicts the CCDF of nodes' degree and PageRank of NVG network for \dsct\ stars. We observe that both nodes' degree and PageRank distributions for HADS stars (red lines) are grouped above levels than the non-HADS (blue lines) stars. The high nodes' degree shows the significant number of connections of nodes with others. However, the high PageRank indicates the important nodes or data points in the light curve with many in-links from other nodes. Since HADS are mostly mono-periodic or double mode stars, so high PageRank nodes (importance nodes) show a significant probability over the other nodes. However, due to the superposition of amplitudes of several modes for non-HADS stars, the high PageRanks (tail of the distribution) have low probability than the low PageRank.  

Analysis for the network properties (e.g., PageRank and nodes' degree distributions) indicates that the V353 Vel, TIC 448892817, and GW Dra stars are in the interface of HADS and non-HADS stars. \cite{LV2022} showed that the TIC 448892817 is a HADS star, confirmed by the present network analysis. Along with the clustering coefficient analysis (Figure \ref{fig3}), the PageRank and degree distribution for RX Cae also suggest considering this star in the HADS subclass. Also, the dominant period (${\rm P}=0.15$ d) and luminosity (${\rm M_V}=1.04$ mag) for RX Cae satisfy the revised P-L relation ${\rm M_V} = (-2.94 \pm 0.06) \log{\rm P} - (1.34 \pm 0.06)$ \citep{ziaali2019}.

\subsection{Effect of the noise on networks' parameters}
\begin{figure}
      \centering
      \begin{subfigure}
         \centering
         \vspace*{-.2cm}
         \includegraphics[width=7.5cm,height=4.5cm]{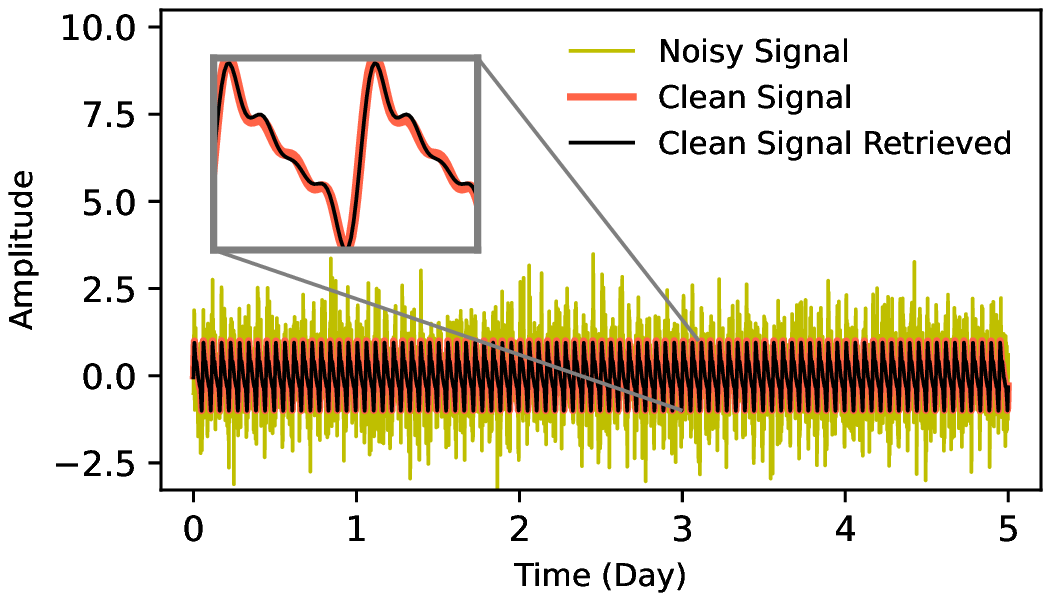}
     \end{subfigure}\\
     \begin{subfigure}
         \centering
         \hspace*{1cm}
         \includegraphics[width=7.5cm,height=4.5cm]{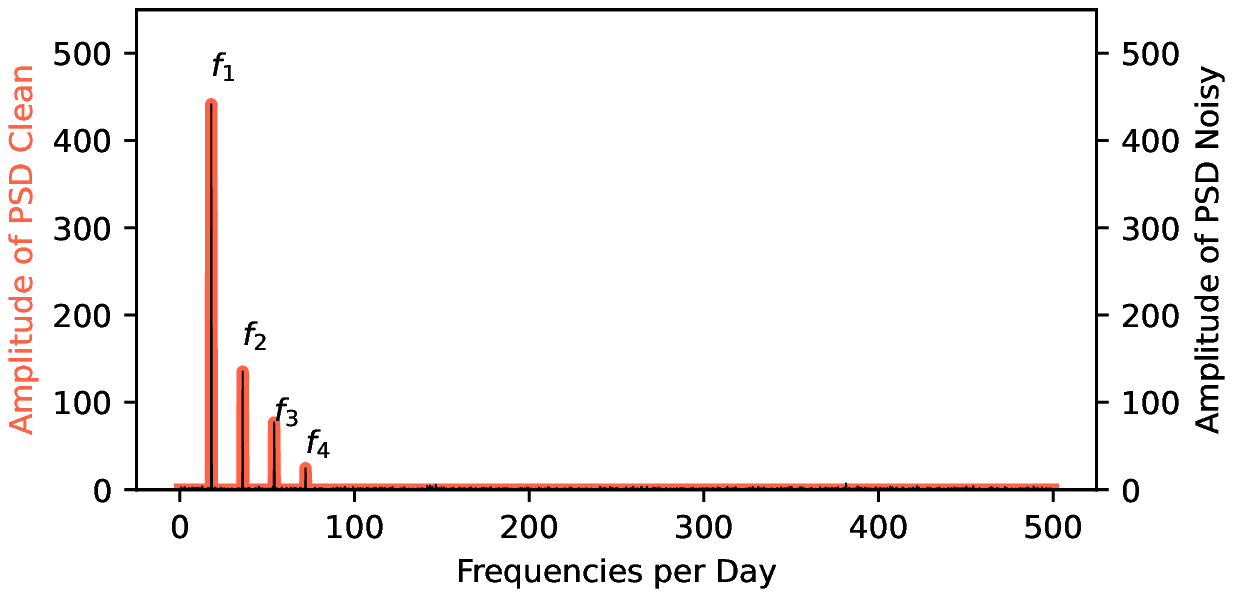}
      \end{subfigure}\\
\caption{Synthetic time series (top panel) and its PSD clean and noisy (bottom panel) signal.}
\label{fig8}
\end{figure}

\begin{figure}
      \centering
      \begin{subfigure}
         \centering
         \vspace*{-.2cm}
         \includegraphics[width=6.5cm,height=4.cm]{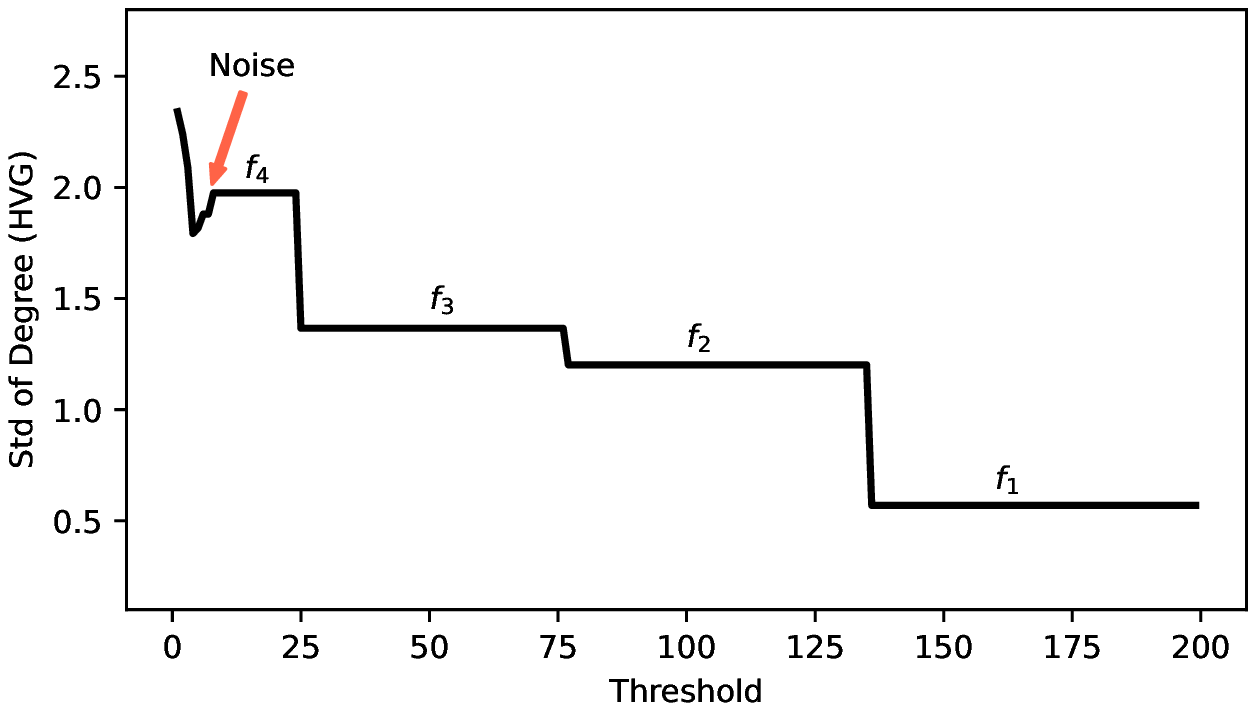}
     \end{subfigure}\\
     \begin{subfigure}
         \centering
         \vspace*{-.3cm}
         \includegraphics[width=6.5cm,height=4.cm]{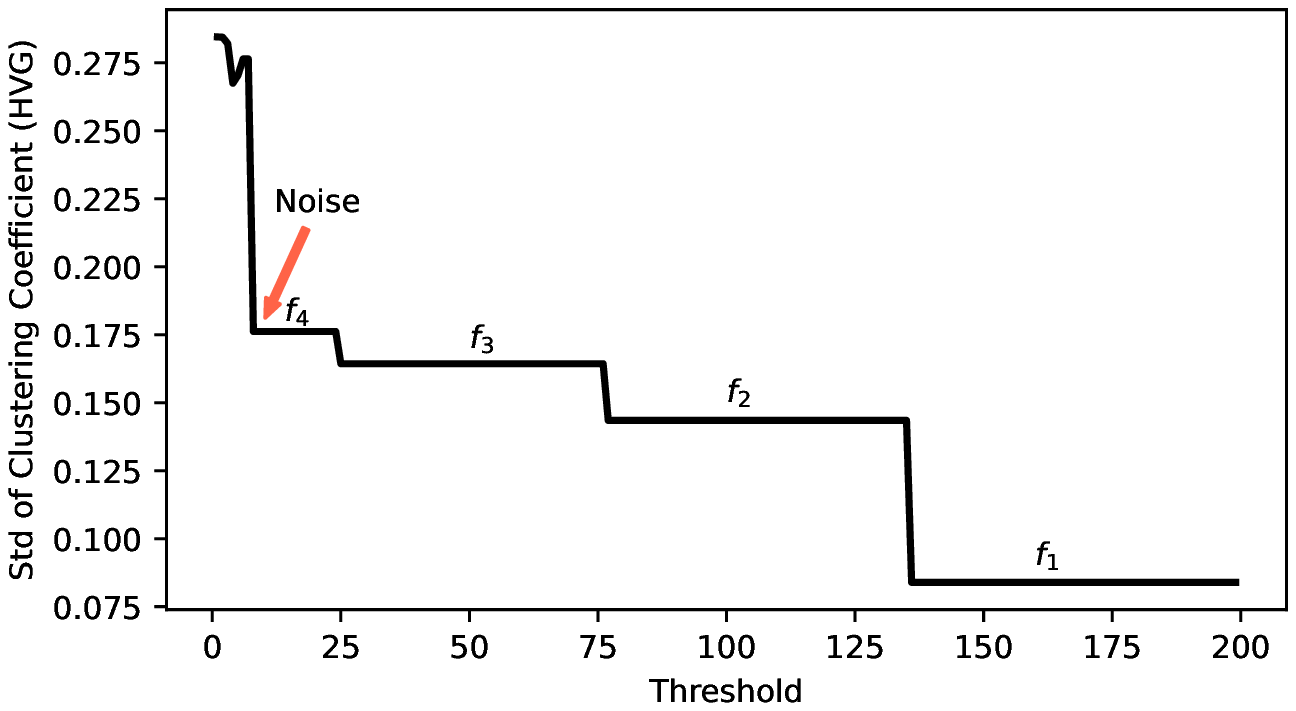}
      \end{subfigure}\\
      \begin{subfigure}
         \centering
         \includegraphics[width=6.5cm,height=4.cm]{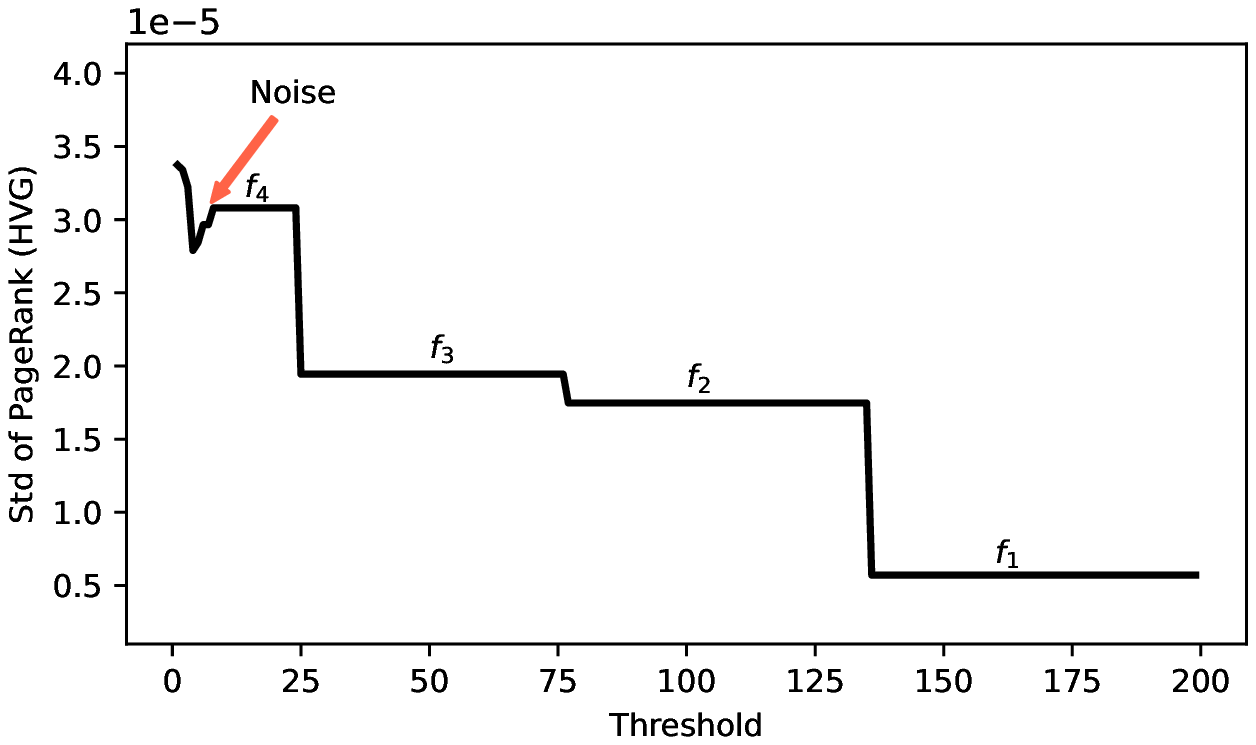}
      \end{subfigure}\\
\caption{Std of network parameters including of nodes degree (top panel), clustering coefficient (middle panel), and PageRank (bottom panel) versus the threshold. Threshed of corresponds to noise and frequencies steps are distinguished.}
\label{fig9}
\end{figure}

    We investigate the effect of noise on the networks' parameters for a synthetic time series with several frequencies and amplitudes (in the range of \dsct\ stars) with harmonics, including the white noises. Figure \ref{fig8} shows the synthetic time series and its power spectral density (PSD) calculated with fast Fourier transformation. We reconstruct the related time series via inverse Fourier transformation for different threshold values as a fraction of PSD subtracted from the primary PSD. For each time series, we build the HVG network. Figure \ref{fig9} displays the standard deviation (std) of network parameters (nodes degree, clustering coefficient, and PageRank) versus the threshold. We observe that the thresholds of decreasing steps for the std of network parameters corresponded to the PSD of frequencies. Therefore, the number of significant steps corresponds to the frequencies made in the time series. Also, the fluctuations of the small thresholds might correspond to the different noise levels. 

\subsection{Reducing the size of the window for light curves} 
Reducing the size of the window of a time series without significantly decreasing the essential characteristics of the time series is very important in machine learning applications \citep{moradkhani2015hybrid}. The modes' information, such as frequencies, amplitudes, and phases, are the main characteristics of pulsating star light curves \citep{aerts2010asteroseismology}. We apply the complex network parameter (e.g., PageRank) to reduce a stellar light curve's data points. To find the key data points of light curves, we first ranked the nodes of a light curve according to their HVG PageRanks. Then we selected a given percentile of data points with higher PageRanks. We extract the characteristics (frequencies and amplitudes) for the resultant light curve. 

\begin{figure}
      \centering
      \begin{subfigure}
         \centering
         \vspace*{-.2cm}
         \includegraphics[width=7.5cm,height=5cm]{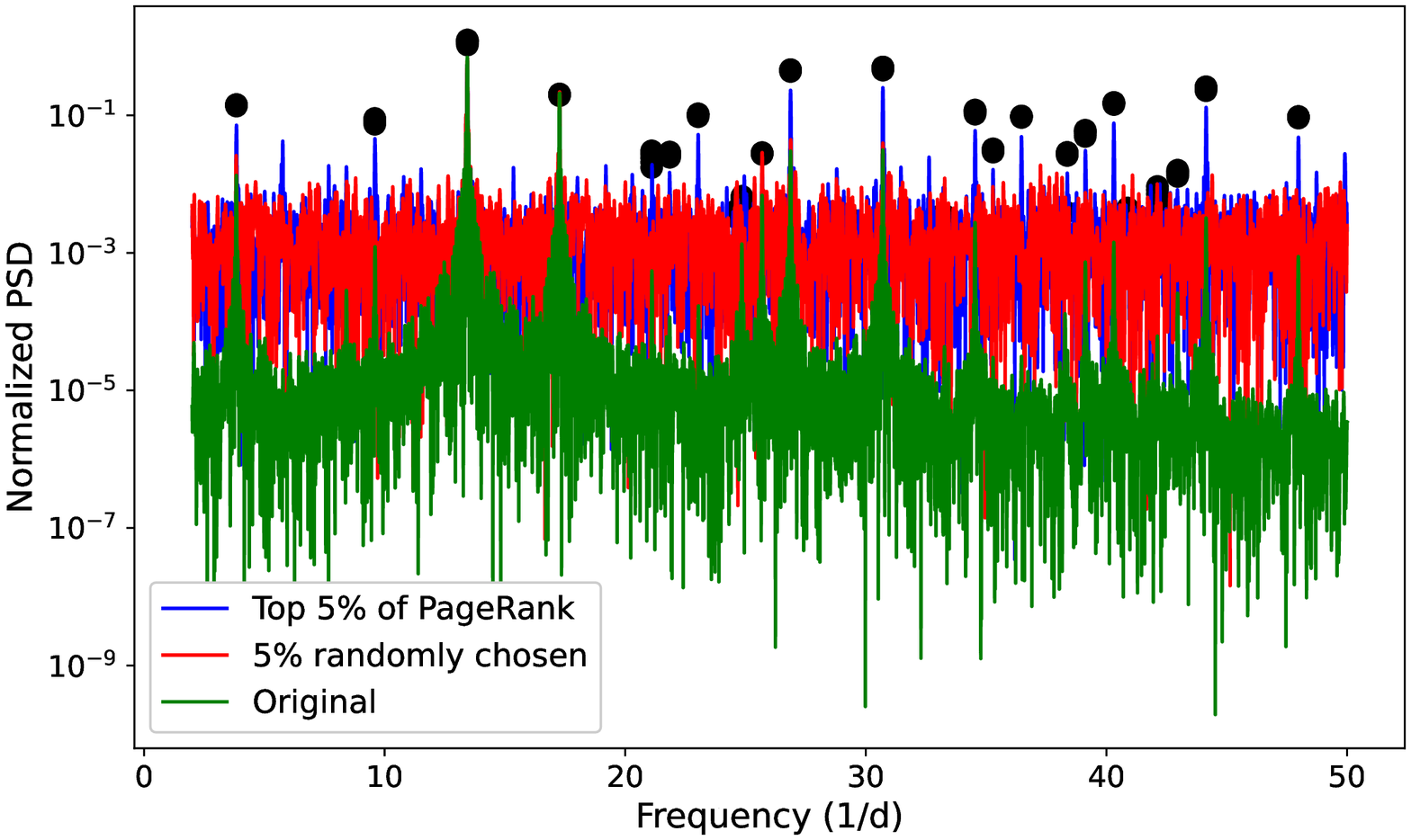}
     \end{subfigure}\\
     \begin{subfigure}
         \centering
         \vspace*{-.3cm}
         \includegraphics[width=7.5cm,height=5cm]{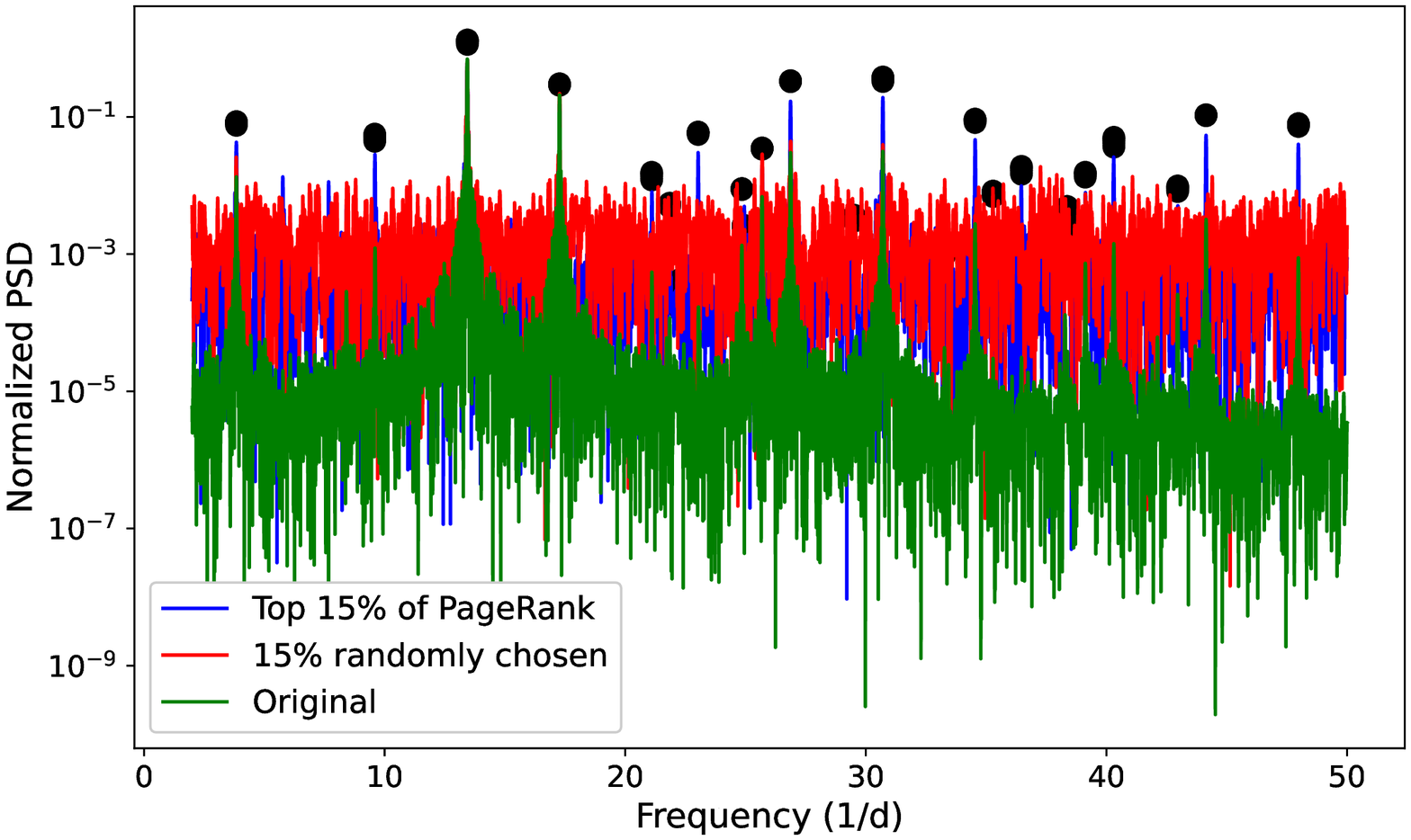}
      \end{subfigure}\\
      \caption{Power spectrum of TIC 448892817 for 5\% (top panel) and 15\%(bottom panel) of light curve selected by the large HVG PageRanks (blue) and randomly chosen (red). The power spectrum of full light curve in sector 5 (green) is added for diagrams. The well-known frequencies are indicated by black circles.}
\label{fig10}
\end{figure}

Figure \ref{fig10} represents the frequency power spectrum for 5\% (top panel) and 15\% (bottom panel) of TIC 448892817 light curve chosen by the large HVG PageRanks (blue) and random sample (red) that compared with the power spectrum of the full light curve in sector 5 (green). We observe that compared with the random sampling, the modes (frequencies, amplitudes, and phases) obtained from light curves reduced by the network approach are significantly satisfied with the full light curves (black circles) reported by \cite{LV2022}. In this case, we also observe that the most amplitudes for identified frequencies were amplified from the noise background. This result implies that reducing the size of the window to e.g., 5\% of the original light curve via the network approach preserves the main characteristics of the light curve. However, some characteristics of the light curve were missed by the random sampling.

\section{Conclusions}\label{concs}
We investigate the network characteristics \dsct\ stars via the HVG and NVG algorithms, which both algorithms study the interaction of data points along the light curves (Figure \ref{fig1}). We focus on the network metrics to discuss the property of HADS and non-HADS stars. 
\begin{itemize}
    \item The linear dependency of the average shortest path length (Figure \ref{fig2}) for HVG network size (in log space) of \dsct\ stars indicates that the networks are small-world complex networks with non-random properties. This small-world property implies that the peaks of the stellar flux light curves connect with some small close peaks and are then linked to other significant peaks along the light curves.       
    \item Scattering of transitivity and average clustering coefficient of HADS and non-HADS stars are clustered in two groups overlapping at some deals (Figure \ref{fig3}). The low average clustering for RX Cae (HD 28837) star is defined as a HADS sample. The lower clustering coefficient for most HADS stars indicates the simple light curve (containing one or two independent modes). In contrast, the higher clustering for non-HADS cases is a signature of the more complicated light curve (including various oscillation modes). These results may be due to HADS ages (low surface gravity) as older than the non-HADS stars.  In other words, the older stars may have simpler light curves than the younger stars.  
    \item The nodes' degree distribution of HVG networks (Figure \ref{fig4} and \ref{fig6}) of \dsct\ stars obey the lognormal distribution. The p-value of K-S statistics rejects both power-law and random network distribution. A lognormal behavior suggests a multiplicative generative mechanism for oscillations of \dsct\ stars.
    \item We find that the nodes' degree and PageRank distributions for NVG networks of HADS and non-HADS stars are grouped in two classes for most cases (Figure \ref{fig7}).
    \item Based on the network ranking for modes, we reduced the window size of a light curve to about 5\% of the original one, which preserves most of the star oscillation modes characteristics (frequencies, amplitudes, and phases). We also observe that most natural modes amplify the amplitude compared to the power spectrum's noise background.  
    
\end{itemize}

Finally, we observe that the network approach can recognize HADS in the \dsct\ stars as a subgroup of pulsating stars. We conclude that the network analysis of light curves helps interpret pulsating stars and their originating mechanisms via stellar dynamics. 
 
\hspace{-0.5cm}
\begin{table*}
\centering 
\caption{Parameters and p-value of fitted lognormal distribution for HADS stars.}
\begin{tabular}{ccc   ccc}
\hline \hline
\multicolumn{3}{c} {Star ID}  & \multicolumn{3}{c}{Lognormal} \\  \cline{1-6} 
\\
Name & HD & TIC & $\mu$ & $\sigma$ & p-value \\  
\cmidrule(l){1-3} \cmidrule(l){4-6} 

AD Cmi&64191  &266328148    &$1.25\pm0.08$& $0.29\pm0.06$   &0.14\\
BE Lyn& 79889 &56914404   &$1.15\pm0.07$& $0.45\pm0.05$   &0.16\\
CY Aqr&--     &422412568  &$1.19\pm0.07$& $0.44\pm0.06$   &0.18\\
DX Cet&16189  &278962831    &$1.35\pm0.06$& $0.33\pm0.06$   &0.14\\
GP And&--     &436546358    &$1.17\pm0.07$& $0.35\pm0.05$   &0.18\\
GW Uma&--     &150276417    &$1.12\pm0.07$& $0.44\pm0.06$   &0.21\\
PT Com&--     &335826251    &$1.19\pm0.07$& $0.38\pm0.06$   &0.28\\           
RS Gru&206379 &139845816    &$1.17\pm0.06$& $0.46\pm0.06$   &0.19\\
V 524 And&--  &196562983 &$1.23\pm0.08$& $0.33\pm0.06$   &0.15\\
V367 Cam&--   &354872568  &$1.27\pm0.06$& $0.34\pm0.04$   &0.19\\
V1162 Ori&-- &34512862 &$1.30\pm0.07$& $0.33\pm0.04$   &0.21\\
V2455 Cyg&204615&266794067 &$1.25\pm0.06$& $0.33\pm0.06$   &0.17\\
YZ Boo&--    &233465540    &$1.21\pm0.07$& $0.37\pm0.05$   &0.16\\
AN Lyn&--    &56882581    &$1.32\pm0.06$& $0.36\pm0.05$   &0.20\\
--&146953& 210548440 &$1.26\pm0.06$& $0.35\pm0.06$   &0.23\\
V1384 Tau&--&415333069 &$1.35\pm0.06$& $0.32\pm0.04$   &0.26\\
V1393 Cen&121517&241787384 &$1.24\pm0.06$& $0.34\pm0.05$   &0.20\\
V2855 Ori&254061 &166979292 &$1.32\pm0.07$& $0.32\pm0.05$   &0.22\\
ZZ Mic&199757&126659093    &$1.35\pm0.06$& $0.27\pm0.03$   &0.06\\
AE Uma&--&357132618    &$1.12\pm0.07$& $0.43\pm0.07$   &0.20\\
$\rho$~ Pup&67523&154360594  &$1.35\pm0.06$& $0.21\pm0.06$   &0.02\\
BL Cam&--    &392774261    &$1.32\pm0.06$& $0.37\pm0.05$   &0.24\\
BS Aqr&223338&9632550    &$1.15\pm0.08$& $0.39\pm0.06$   &0.17\\
DE Lac&--    &119486942    &$1.26\pm0.07$& $0.38\pm0.05$   &0.27\\
EH Lib&--    &157861023    &$1.04\pm0.08$& $0.42\pm0.06$   &0.20\\
KZ Hya&94033 &188209486    &$1.15\pm0.07$& $0.39\pm0.06$   &0.19\\
RY Lep&38882 &93441696     &$1.28\pm0.04$& $0.29\pm0.02$   &0.08\\
SS Psc&--&456857185     &$1.36\pm0.06$& $0.33\pm0.05$   &0.32\\
SX Phe&223065&224285325     &$1.26\pm0.06$& $0.35\pm0.06$   &0.20\\
V1719 Cyg&200925&290277380  &$1.25\pm0.04$& $0.30\pm0.03$   &0.30\\
VX Hya&--&289711518     &$1.29\pm0.05$& $0.40\pm0.06$   &0.30\\
VZ Cnc&73857&366632312     &$1.17\pm0.06$& $0.52\pm0.06$   &0.21\\
XX Cyg&--    &233310793    &$1.07\pm0.07$& $0.45\pm0.07$   &0.29\\

\hline \hline
\end{tabular}
\label{tab1}
\end{table*}

\hspace{-0.5cm}

\begin{table}
\caption{ Parameters and p-value of fitted lognormal distribution for non-HADS stars.}
\centering 
\begin{tabular}{ccc   ccc}
\hline \hline
\multicolumn{3}{c} {Star ID}  & \multicolumn{3}{c}{Lognormal} \\  \cline{1-6} 
\\
Name & HD & TIC & $\mu$ & $\sigma$ & p-value \\  
\cmidrule(l){1-3} \cmidrule(l){4-6} 
  
--& 129831& 81003     &$0.93\pm0.09$& $0.47\pm0.07$  &0.35\\
--&77914& 975071    &$0.92\pm0.09$& $0.47\pm0.07$  &0.29\\
--&112063& 9591460   &$0.89\pm0.09$& $0.48\pm0.07$  &0.29\\
RX Cae&28837& 7808834   &$1.11\pm0.07$& $0.34\pm0.05$  &0.19\\
CV Phe&13755& 7245720   &$1.17\pm0.07$& $0.35\pm0.05$  &0.32\\
--&21295& 12524129  &$0.92\pm0.09$& $0.58\pm0.07$  &0.38\\
--&79111& 18658256  &$0.88\pm0.08$& $0.48\pm0.06$  &0.35\\
--&86312& 26957587  &$0.91\pm0.08$& $0.52\pm0.07$  &0.35\\
--&181280& 30624832  &$0.89\pm0.09$& $0.57\pm0.07$  &0.38\\
--&44930& 34737955  &$1.16\pm0.07$& $0.45\pm0.07$  &0.40\\
--&185729& 79659787  &$0.91\pm0.09$& $0.50\pm0.08$  &0.33\\
--&25674& 34197596  &$1.28\pm0.07$& $0.41\pm0.06$  &0.24\\
--&113221& 102192161 &$0.97\pm0.08$& $0.57\pm0.08$  &0.42\\
--&180349& 121729614 &$0.97\pm0.09$& $0.57\pm0.09$  &0.43\\
--&216728& 137796620 &$0.91\pm0.09$& $0.59\pm0.08$  &0.38\\
--&113211& 253296458 &$0.92\pm0.08$& $0.58\pm0.08$  &0.43\\
--&31322& 246902545 &$1.31\pm0.06$& $0.37\pm0.05$  &0.28\\
--&32433& 348792358 &$1.03\pm0.08$& $0.48\pm0.07$  &0.38\\
--&56843& 387235455 &$0.97\pm0.08$& $0.50\pm0.07$  &0.38\\
--&38597& 100531058 &$1.22\pm0.07$& $0.41\pm0.06$  &0.25\\
--&--& 448892817 &$0.88\pm0.08$& $0.49\pm0.07$  &0.29\\
--&99302& 458689740 &$0.97\pm0.08$& $0.50\pm0.07$  &0.41\\
V479 Tau&24550& 459908110 &$0.98\pm0.08$& $0.51\pm0.07$  &0.39\\
V353 Vel&93298& 106886169 &$1.21\pm0.07$& $0.38\pm0.06$  &0.29\\
--&17341& 122615966 &$1.26\pm0.06$& $0.41\pm0.06$  &0.29\\
--&183281& 137341551 &$1.13\pm0.07$& $0.51\pm0.07$  &0.32\\
--&182895& 159647185 &$1.37\pm0.07$& $0.42\pm0.06$  &0.36\\
--&46722& 172193026 &$1.28\pm0.06$& $0.34\pm0.04$  &0.24\\
--&8043& 196921106 &$1.3\pm0.06$ & $0.46\pm0.03$  &0.38\\
CC Gru&214441& 161172103 &$0.87\pm0.06$& $0.45\pm0.06$  &0.16\\
--&24572& 242944780 &$1.32\pm0.06$& $0.41\pm0.07$  &0.31\\
--&--& 274038922 &$1.37\pm0.06$& $0.34\pm0.05$  &0.29\\
IN Dra&191804& 269697721 &$1.29\pm0.06$& $0.35\pm0.05$  &0.31\\
GW Dra&--& 329153513 &$1.31\pm0.05$& $0.35\pm0.05$  &0.27\\
CP Oct&21190& 348772511 &$1.32\pm0.06$& $0.33\pm0.05$  &0.27\\
IO Dra&193138& 403114672 &$1.34\pm0.06$& $0.34\pm0.05$  &0.31\\
--&42005& 408906554 &$1.17\pm0.07$& $0.46\pm0.05$  &0.33\\
DE Cmi&67852& 452982723 &$1.32\pm0.06$& $0.34\pm0.04$  &0.28\\
V435 Car&44958&255548143      &$1.32\pm0.06$& $0.35\pm0.05$  &0.32\\
V1790 Ori&290799&11361473     &$1.23\pm0.07$& $0.45\pm0.06$  &0.33\\
\hline \hline
\end{tabular}
\label{tab2}
\end{table}

\clearpage
\bibliographystyle{apalike}

\bibliography{bibtex.bib}


\end{document}